\documentclass[12pt]{article}
\usepackage{fancyhdr}

\usepackage{graphicx}
\graphicspath{ {./images/} }
\usepackage[dvipsnames]{xcolor}
\usepackage{mathrsfs}
\usepackage[T1]{fontenc}
\usepackage{mathpazo}
\usepackage{setspace}
\usepackage{amsfonts}
\usepackage{amssymb}
\usepackage{amsmath}
\usepackage{epsfig}
\usepackage{latexsym}
\usepackage{color}
\usepackage{graphicx}
\usepackage{nicefrac}
\usepackage[latin1]{inputenc}
\usepackage{slashed}
\usepackage{multirow}
\usepackage{comment}
\usepackage{soul}
\usepackage{hyperref}
\usepackage{datetime}
\usepackage{colortbl}

\usepackage[titletoc]{appendix}

\def\hybrid{\topmargin -20pt    \oddsidemargin 0pt
	\headheight 0pt \headsep 0pt
	\textwidth 6.25in       
	\textheight 9.25in       
	\marginparwidth .875in
	\parskip 5pt plus 1pt   \jot = 1.5ex}

\hybrid

\def\baselinestretch{1.2}

\catcode`\@=11

\def\marginnote#1{}
\newcount\hour
\newcount\minute
\newtoks\amorpm
\hour=\time\divide\hour by60
\minute=\time{\multiply\hour by60 \global\advance\minute by-\hour}
\edef\standardtime{{\ifnum\hour<12 \global\amorpm={am}%
		\else\global\amorpm={pm}\advance\hour by-12 \fi
		\ifnum\hour=0 \hour=12 \fi
		\number\hour:\ifnum\minute<10 0\fi\number\minute\the\amorpm}}
\edef\militarytime{\number\hour:\ifnum\minute<10 0\fi\number\minute}

\def\draftlabel#1{{\@bsphack\if@filesw {\let\thepage\relax
			\xdef\@gtempa{\write\@auxout{\string
					\newlabel{#1}{{\@currentlabel}{\thepage}}}}}\@gtempa
		\if@nobreak \ifvmode\nobreak\fi\fi\fi\@esphack}
	\gdef\@eqnlabel{#1}}
\def\@eqnlabel{}
\def\@vacuum{}
\def\draftmarginnote#1{\marginpar{\raggedright\scriptsize\tt#1}}

\def\draft{\oddsidemargin -.5truein
	\def\@oddfoot{\sl preliminary draft \hfil
		\rm\thepage\hfil\sl\today\quad\militarytime}
	\let\@evenfoot\@oddfoot \overfullrule 3pt
	\let\label=\draftlabel
	\let\marginnote=\draftmarginnote
	\def\@eqnnum{(\theequation)\rlap{\kern\marginparsep\tt\@eqnlabel}%
		\global\let\@eqnlabel\@vacuum}  }

\def\preprint{\twocolumn\sloppy\flushbottom\parindent 2em
	\leftmargini 2em\leftmarginv .5em\leftmarginvi .5em
	\oddsidemargin -.5in    \evensidemargin -.5in
	\columnsep .4in \footheight 0pt
	\textwidth 10.in        \topmargin  -.4in
	\headheight 12pt \topskip .4in
	\textheight 6.9in \footskip 0pt
	\def\@oddhead{\thepage\hfil\addtocounter{page}{1}\thepage}
	\let\@evenhead\@oddhead \def\@oddfoot{} \def\@evenfoot{} }

\def\numberbysection{\@addtoreset{equation}{section}
	\def\theequation{\thesection.\arabic{equation}}}

\def\underline#1{\relax\ifmmode\@@underline#1\else
	$\@@underline{\hbox{#1}}$\relax\fi}

\def\titlepage{\@restonecolfalse\if@twocolumn\@restonecoltrue\onecolumn
	\else \newpage \fi \thispagestyle{empty}\c@page\z@
	\def\thefootnote{\fnsymbol{footnote}} }

\def\endtitlepage{\if@restonecol\twocolumn \else \newpage \fi
	\def\thefootnote{\arabic{footnote}}
	\setcounter{footnote}{0}}  

\catcode`@=12
\relax

\def\figcap{\section*{Figure Captions\markboth
		{FIGURECAPTIONS}{FIGURECAPTIONS}}\list
	{Figure \arabic{enumi}:\hfill}{\settowidth\labelwidth{Figure
			999:}
		\leftmargin\labelwidth
		\advance\leftmargin\labelsep\usecounter{enumi}}}
 \relax
\def\tablecap{\section*{Table Captions\markboth
		{TABLECAPTIONS}{TABLECAPTIONS}}\list
	{Table \arabic{enumi}:\hfill}{\settowidth\labelwidth{Table
			999:}
		\leftmargin\labelwidth
		\advance\leftmargin\labelsep\usecounter{enumi}}}
 \relax
\def\reflist{\section*{References\markboth
		{REFLIST}{REFLIST}}\list
	{[\arabic{enumi}]\hfill}{\settowidth\labelwidth{[999]}
		\leftmargin\labelwidth
		\advance\leftmargin\labelsep\usecounter{enumi}}}
 \relax
\makeatletter
\newcounter{pubctr}
\def\publist{\@ifnextchar[{\@publist}{\@@publist}}
\def\@publist[#1]{\list
	{[\arabic{pubctr}]\hfill}{\settowidth\labelwidth{[999]}
		\leftmargin\labelwidth
		\advance\leftmargin\labelsep
		\@nmbrlisttrue\def\@listctr{pubctr}
		\setcounter{pubctr}{#1}\addtocounter{pubctr}{-1}}}
\def\@@publist{\list
	{[\arabic{pubctr}]\hfill}{\settowidth\labelwidth{[999]}
		\leftmargin\labelwidth
		\advance\leftmargin\labelsep
		\@nmbrlisttrue\def\@listctr{pubctr}}}
 \relax
\makeatother
\newskip\humongous \humongous=0pt plus 1000pt minus 1000pt

\newif\ifdtup

\relax

\def\be{\begin{equation}}
\def\ee{\end{equation}}
\def\ba{\begin{eqnarray}}
\def\ea{\end{eqnarray}}

\def\del{\partial}

\def\b{\beta}

\def\d{\delta}

\def\e{\epsilon}

\def\p{\pi}

\def\m{\mu}
\def\n{\nu}

\def\l{\lambda}

\def\s{\sigma}

\def\no{\noindent}

\def\qq{\qquad}

\def\IR{\relax{\rm I\kern-.18em R}}

\def \ov {\over}

\def\IR{\relax{\rm I\kern-.18em R}}
\def\IL{\relax{\rm I\kern-.18em L}}

\def\inv{^{\raise.15ex\hbox{${\scriptscriptstyle -}$}\kern-.05em 1}}

\def\tr{{\rm tr}}
\def\Tr{{\rm Tr}}

\begin{document}
\renewcommand{\theequation}{\thesection.\arabic{equation}}
\csname @addtoreset\endcsname{equation}{section}

\newcommand{\beq}{\begin{equation}}
\newcommand{\eeq}[1]{\label{#1}\end{equation}}
\newcommand{\ber}{\begin{equation}}
\newcommand{\eer}[1]{\label{#1}\end{equation}}
\newcommand{\eqn}[1]{(\ref{#1})}
\begin{titlepage}
	\begin{center}
		

		${}$
		\vskip .2 in

		{\large\bf Asymmetric CFTs arising at the IR fixed points of RG flows}
		
		\vskip .2 in


		\vskip 0.4in
		
		{\bf George Georgiou},\ \ {\bf Georgios P. D. Pappas}\ \ and\ \ {\bf Konstantinos Sfetsos}
		\vskip 0.1in

		{\em
			Department of Nuclear and Particle Physics,\\
			Faculty of Physics, National and Kapodistrian University of Athens,\\
			15784 Athens, Greece\\
		}

		\vskip 0.1in

		{ \footnotesize{george.georgiou@phys.uoa.gr, geopappas@phys.uoa.gr, ksfetsos@phys.uoa.gr}}
		
		
		\vskip .5in
	\end{center}
	
	\centerline{\bf Abstract}

	\no
	We construct a generalization of the cyclic $\l$-deformed models of \cite{Georgiou:2017oly} by relaxing the requirement that all the WZW models should  have the same level $k$.  
	Our theories are integrable and flow from a single UV point to different IR fixed points depending on the different orderings of the WZW levels $k_i$. 
	First we calculate the Zamolodchikov's C-function for these models as exact functions of the deformation parameters. 
	Subsequently, we fully characterize each of the IR conformal field theories. 
Although the corresponding  left and right sectors have different symmetries, realized as products of current and coset-type symmetries,   the associated central charges are precisely equal, in agreement with the values
obtained from the C-function.

	\vskip .4in
	\noindent
\end{titlepage}
\vfill
\eject

\newpage

\tableofcontents

\noindent

\def\baselinestretch{1.2}
\baselineskip 20 pt
\noindent


\setcounter{equation}{0}

\section{Introduction}
\label{Introduction}

It is an extremely important and interesting endeavour to obtain exact results for quantum field theories (QFTs) living in any number of dimensions.
The importance lies on the fact that based on this achievement one gains access to the non-perturbative regime of the theory, on one hand, and may uncover hidden symmetries that are not at all apparent from the Lagrangian formulation of the theory, on the other. Recently, this goal was achieved  for a class of two-dimensional QFTs. These theories can be formulated as current-current perturbations of conformal field theories (CFTs) of the WZW type.

The systematic construction of a large class of integrable two-dimensional field theories based on group spaces $G$ and having an explicit Lagrangian formulation was performed in \cite{Sfetsos:2013wia,Georgiou:2016urf,Georgiou:2017jfi,Georgiou:2018hpd,Georgiou:2018gpe,Driezen:2019ykp}. The quantum properties of these theories were further studied in 
\cite{Georgiou:2015nka,Georgiou:2016iom,Georgiou:2016zyo,Georgiou:2017oly,Itsios:2014lca,Georgiou:2017aei}. 
These models are parametrized by several couplings, for small values of which the models take the form of  one or more WZW models \cite{Witten:1983ar}  perturbed by current bilinears. In the context of these models a large class of observables, including $\beta$-functions\cite{Kutasov:1989dt,Gerganov:2000mt,Itsios:2014lca,Sfetsos:2014jfa,Appadu:2015nfa}, anomalous dimensions of currents and primary operators \cite{Georgiou:2015nka,Georgiou:2016iom,Georgiou:2016zyo,Georgiou:2019jcf,Georgiou:2019aon}, three-point correlators of currents and/or primary fields \cite{Georgiou:2015nka,Georgiou:2019aon} were computed.
In addition, Zamolodchikov's C-function \cite{Zamolodchikov:1986gt} in these models were calculated 
as {\it exact} functions of the deformation parameters \cite{Georgiou:2018vbb,Sagkrioti:2018abh}.\footnote{These results are exact in the deformation parameters but leading in the $1/k$-expansion. Recently, the subleading terms in this expansion were obtained in 
\cite{Georgiou:2019nbz,Hoare:2019mcc} for the $\beta$-functions, the C-function and the anomalous dimensions of the operators perturbing the CFT in the cases of group and coset spaces.} This goal was achieved by using a variety of complementary methods. One way \cite{Georgiou:2016iom,Georgiou:2015nka} was to use low order perturbation theory around the conformal point in conjunction with certain non-perturbative  symmetries \cite{Kutasov:1989aw} in the space of couplings which these theories generically exhibit. A second method is based on the geometry in the space of couplings assisted by the knowledge of the all-loop effective action of these models. This method  allows in principle the calculation of the anomalous dimensions of composite operators built from an arbitrary number of currents was developed in \cite{Georgiou:2019jcf}. More recently, yet another method for calculating exact results in this class of models was initiated in \cite{Georgiou:2019aon}. This method was very recently applied to deformed coset CFTs in
\cite{Georgiou:2020bpx}. In this method  one ends up performing calculations around the free field point and not around the conformal one, which is much easier and in addition all 
deformation effects are captured by the couplings in the interaction vertices.

The virtue of the models constructed in \cite{Georgiou:2017jfi,Georgiou:2018hpd,Georgiou:2018gpe} for deformations
based on current algebras  and in \cite{Sfetsos:2017sep} for deformations cases of coset CFTs, as compared to the prototype $\l$-deformed model \cite{Sfetsos:2013wia} (for $SU(2)$ the $\l$-deformed model was found earlier in \cite{Balog:1993es}) is that they possess a rich structure in their RG flow consisting of several fixed points with a different CFT sitting at each fixed point. It remains an open problem to fully classify these CFTs by determining their symmetry groups. 
 In this work we make a major advancement towards this direction. In particular, we will firstly construct a generalization of the cyclic model presented in \cite{Georgiou:2017oly} by relaxing the requirement that all the WZW models employed in the construction have the same level $k$.  In this way, we will obtain a class of models 
which under the renormalization group (RG) flow from the single 
ultraviolet (UV) point to different infrared (IR) fixed points depending on the different orderings of the WZW levels $k_i$.  
This was not the case in the original model of 
\cite{Georgiou:2017oly} which did not have any IR fixed points due to the equality of the levels. 
Subsequently, we will calculate the C-functions of these models as exact functions of the deformation parameters. This will be the first step in the complete characterization of
each of the conformal field theories that live at each of the IR fixed points. 

Our result consists of  the remarkable fact that although the left and right sectors of the corresponding IR CFTs have different symmetries, which are products of current and coset-type symmetries (see tables \ref{table:eee'} and \ref{table:ee'}), 
the central charge of the left and right sectors are precisely equal, in agreement with the expressions for the C-functions which do not distinguish the left from the right sector.  We note that left-right asymmetric coset CFTs with equal left-right central charges were realized before in \cite{Thomas} as CFTs corresponding to the left-right asymmetric gauged actions of 
\cite{2,Bars:1991pt}. Finally, let us mention that the models presented in this
work are integrable but we do not make any direct use of this fact in our considerations.

The paper is organized as follows: In section 2, we will firstly review the model of \cite{Georgiou:2017oly} and then we will present a certain generalization of it suited for our purposes.   In section 3.1, we will calculate the, exact in the deformation parameters $\l_i$, C-function of the aforementioned theory from which we will read off the central charge of the IR CFTs. Its form will give us a strong indication for the type of symmetries of the IR CFTs. The emerging symmetries at the IR will be  products of chiral current and coset-type symmetries. In section 3.2, we will present a method for identifying the chiral chiral symmetries in the case of two group elements, $n=2$. This will be based on  the identification of the chiral currents generating the left-right group transformations and the calculation of the Poisson brackets (PBs) for them using the canonical formalism. This method for identifying the chiral chiral symmetries  is easily generalized to the case of arbitrary $n$. In section 3.3, we will investigate the type of the 
IR fixed point CFTs at the IR in all detail by developing a formalism which will give us access, also to the coset-type part of the conformal symmetry. We will present it in detail for $n=2,3$ and $n=4$. This formalism will render us capable of identifying the type of the CFTs in the IR regime for arbitrary $n$. 
We present our conclusions in section 4. Finally, in appendix A we prove that our models are integrable by finding the appropriate Lax pairs, and in appendix B we collect some formulae regarding the Hamiltonian formulation of our $n=2$ model.

\section{Constructing closed chain models}
\label{models}

In this section, we fist briefly review the models constructed in \cite{Georgiou:2017oly} which will serve as a basis in  
subsequent developments in this paper. 
These models represent the effective action of coupled WZW models all at the same level $k$ with the characteristic that when the deviation from the  conformal point is small they interact via current bilinears forming a closed chain. 
Then we will generalize this construction to the case of arbitrary levels for the corresponding WZW models. As we will see having different levels warranties that there will be an IR fixed point under the RG flow, as well as the avoidance of the strong coupling regime. 

\no
The basic idea \cite{Georgiou:2016urf,Georgiou:2017oly} was to 
start with the sum of $n$ WZW models all at level $k$, based on the same semi-simple group $G$ to which one adds  $n$ PCM models with non-isotropic, in general, coupling matrices. Next, one gauges their common global symmetry
\begin{align}
\label{a}
g_i&\to h^{-1}_ig_i h_{i+1}\, ,\quad \tilde{g_i}\to h_i\tilde{g_i}\,,\quad i=1,2,\dots , n\,,
\end{align}
where $g_i\in G$ refers to each of the WZW, $\tilde{g}_i \in G$ refers to the PCM models  and $h_i \in G$ denotes the local symmetry transformation. This kind of gauging resembles the one used for gauged WZW in \cite{2,Bars:1991pt}. 
The index $i$ is defined modulo $n$. Fixing the gauge by choosing  $\tilde{g_i}=\mathbb{1}\ , \forall\ i$ gives the action \cite{Georgiou:2017oly}
\begin{equation}
\label{b}
\begin{split}
S=&\sum_{i=1}^{n}S_{k}(g_i)+\frac{k}{\pi}\int d^2\s\sum_{i=1}^{n}\Tr\Big(A^{(i)}_-\partial_+g_ig_i^{-1}-A^{(i+1)}_+g_i^{-1}\partial_-g_i+\\
&+A^{(i)}_-g_iA^{(i+1)}_+g_i^{-1}-A^{(i)}_+\l_i^{-1}A^{(i)}_-\Big)\,,
\end{split}
\end{equation}
where $\l_i$ denote arbitrary matrices while $S_k(g_i)$ represents the WZW action for the field $g_i$, at level $k$. Finally, the gauge fields $A^{(i)}_{\pm}$ belong in the algebra $\mathcal{L}(G)$ of $G$. We emphasize that the whole gauging procedure
is such that  that the resulting action is free of gauge anomalies \cite{Georgiou:2017oly}.

The effective action that is obtained after we integrate out in  \eqref{b} the gauge fields $A^{(i)}_{\pm}$ takes, for small values of the matrices $\l_i$, the form of  $n$ distinct WZW models interacting by mutual current bilinears
\begin{equation}
\label{c}
S=\sum_{i=1}^{n}S_k(g_i)+\frac{k}{\pi}\int d^2\s\sum_{i=1}^{n}\Tr\Big(J^{(i+1)}_+\l_{i+1}J^{(i)}_- +\mathcal{O}(\l^2)\Big)\, ,
\end{equation}
where 
\be
J^{(i)}_+= -i \del_+ g_i g_i^{-1}\, , \quad J^{(i)}_-= -i g_i^{-1}\del_- g_i\, ,\quad  i=1,2,\dots , n\ .
\ee
We see that when the strength of the interactions, controlled  by the magnitude of the elements of the matrices $\l_i$, 
is small, the interactions are bilinear in the current of the nearest neighbors  and in such a way that a closed chain is formed. 
This property, together with the fact that all the WZW models have the same level, warrants an anomaly free gauging procedure.
The exact in $\lambda_i$ effective action can be found in \cite{Georgiou:2017oly} and will not be needed for our purposes. 
In addition, a Hamiltonian analysis of the action \eqref{b} revealed that the model at hand is canonically equivalent to $n$ independent single $\l$-deformed WZW-models each with a coupling $\l_i$. Thus the RG flow equations of 
each of the coupling matrices $\l_i$ should be the same  as that of a single $\l$-deformed model \cite{Georgiou:2017oly}.
This fact may also be seen from the form of the interaction terms \eqn{c} and conformal perturbation theory arguments,
as discussed in \cite{Georgiou:2017oly} as well. 

We will study a generalization of \eqref{b} in which the levels $k_i$ of the  WZW models will be different for adjacent sites. Then, the 
analog of the action \eqn{b} is
\begin{equation}
\label{d}
\begin{split}
S=&\sum_{i=1}^{n}S_{k_i}(g_i)+{1\ov \pi} \int d^2\s \sum_{i=1}^{n} k_i \Tr\big(A^{(i)}_-\partial_+g_ig_i^{-1}-A^{(i+1)}_+g_i^{-1}\partial_-g_i 
\\
&+A^{(i)}_-g_iA^{(i+1)}_+g_i^{-1}\big)- k^{(i)}\Tr\big(A^{(i)}_+\l_i^{-1}A^{(i)}_-\big)\, .
\end{split}
\end{equation}
Solving the equations of motion of $A^{(i)}_{\pm}$ we find that
\begin{equation}
\label{2.5}
A^{(1)}_+=i(\mathbb{1}-\hat{D}_1\dots\hat{D}_n)^{-1}\sum_{i=1}^{n}\hat{D}_1\dots\hat{D}_{i-1}(\l^{(i)}_0)^{-1}\l^T_iJ^{(i)}_+\,.
\end{equation}
In the above equations the various parameters are defined as
\begin{equation}
\label{kli}
k^{(i)}=\sqrt{k_ik_{i-1}}\, ,\quad\l_0^{(i)}=\sqrt{\frac{k_{i-1}}{k_i}}\, ,\quad i=1,2,\dots,n \ .
\end{equation}
Note that, not all $\l_0$'s are independent since they obey the identity $\prod_{i=1}^n \l_0^{(i)} = 1$. In addition
\be
\hat{D}_i=(\l^{(i)}_0)^{-1}\l_i^TD_i\, ,\quad  (D_i)_{ab} = \Tr(t_a g_i t_b g_i^{-1})\, ,\quad  i=1,2,\dots , n\ ,
\ee
for a set of representation matrices $t_a$, $a=1,2,\dots,\dim G$, normalized to unity and obeying the corresponding Lie algebra.
The rest of the fields are obtained by cyclic permutations.\footnote{For example 
	\begin{equation}\label{9}
	A^{(2)}_+=i(\mathbb{1}-\hat{D}_2\dots\hat{D}_n\hat{D}_1)^{-1}\sum_{i=2}^{n+1}\hat{D}_2\dots\hat{D}_{i-1}(\l_0^{(i)})^{-1}\l_i^TJ^{i}_+\,.
	\end{equation}}
Note that, the sequence of the $\hat{D}_i$ operators that appears in the sum should be strictly increasing, otherwise it should be 
replaced by the identity matrix. 

\no
The action \eqn{d} is permutation invariant in the index $i$, as well as invariant under the generalized parity transformation 
\begin{equation}
\label{sysmmmr}
\s_+\leftrightarrow\s_-\, ,\quad g_i\to g^{-1}_{n+2-i}\, ,\quad k_i\to k_{n+2-i}\, , \quad 
 \l_i\to \l^T_{n+3-i}\, ,\quad  A_\pm^{(i)}\to A_\mp^{(n+3-i)} \ .
\end{equation}
Using this and  \eqn{2.5} we may find the on-shell values of the gauge fields $A_-^{(i)}$ as well. 

\no
Plugging the gauge fields into \eqn{d} we find the $\s$-model action\footnote{
Cyclic in the indices $1,2,\dots, n$ below means that the next term is
\begin{equation}
\Tr\bigg(\frac{1}{2}J^{(2)}_+D_2\frac{\mathbb{1}+\hat{D}^T_2\hat{D}^T_1\dots\hat{D}^T_3}{\mathbb{1}-\hat{D}^T_2\hat{D}^T_1\dots\hat{D}^T_3}J^{(2)}_-+\sum_{i=3}^{n+1}J^{(i)}_+(\l^{(i)}_0)^{-1}\l_i\hat{D}^T_{i-1}\dots \hat{D}^T_3(\mathbb{1}-\hat{D}^T_2\hat{D}^T_1\dots \hat{D}^T_3)^{-1}J^{(2)}_-\bigg)
\end{equation}
}
\ba
\label{10}
& &S=\frac{k_1}{12\pi}\int\tr(g_1^{-1}dg_1)^3+\frac{k_1}{\pi}\int d^2\s\Tr\bigg(\frac{1}{2}J^{(1)}_+D_1\frac{\mathbb{1}+\hat{D}^T_1\hat{D}^T_n\dots\hat{D}^T_2}{\mathbb{1}-\hat{D}^T_1\hat{D}^T_n\dots\hat{D}^T_2}J^{(1)}_-
\\
&&+\sum_{i=2}^{n}J^{(i)}_+(\l^{(i)}_0)^{-1}\l_i\hat{D}^T_{i-1}\dots \hat{D}^T_2(\mathbb{1}-\hat{D}^T_1\hat{D}^T_n\dots \hat{D}^T_2)^{-1}J^{(1)}_-\bigg)+ \text{(cyclic in $1,2,\dots,n$)}\,.
\nonumber
\ea
This action is of course invariant under  the generalized parity transformation  \eqn{sysmmmr}. In addition,
for small values of the coupling matrices $\l_i$ we have that
\begin{align}
\label{d1}
S&=\sum_{i=1}^{n}S_{k_i}(g_i)+\sum_{i=1}^{n}k^{(i+1)}\int d^2\s \Tr(J^{(i+1)}_+\l_{i+1}J^{(i)}_-)+\mathcal{O}(\l^2)\,.
\end{align}
For diagonal matrices $\l_i$ the above $\s$-model action \eqn{10} is integrable and this is explicitly demonstrated Appendix A. 
The Hamiltonian analysis for two-coupled models was performed in  \cite{Georgiou:2017jfi} for the equal level case.
The extension of this for unequal levels  can be found in Appendix B. 

\no
As in the case of equal levels, the RG flow equations for the matrices $\l_i$ are decoupled.
This is apparent from the form of the interactions in \eqn{d1} since different terms have trivial OPE's among 
themselves.\footnote{For general matrices the RG flow equations can be found  for the equal level case in \cite{Sfetsos:2014jfa} and for unequal levels in  \cite{Sagkrioti:2018rwg}. }
For the purposes of the present paper we consider the isotropic case in which $(\l_i)_{ab} = \l_i \d_{ab}$. 
Then,  the $\b$-function for each of the couplings $\l_i$ reads 
\begin{equation}\label{e}
\frac{d\l_i}{d \ln \m^2}=-\frac{c_G}{2k^{(i)}}\frac{\l_i^2(\l_i-\l^{(i)}_0)(\l_i-(\l^{(i)}_0)^{-1})}{(1-\l_i^2)^2}\,, \quad
i=1,2,\dots , n \ ,
\end{equation}
where $\m$ is the renormalization energy scale. As we can see the running of each of the deformation parameters does not depend on the other couplings but only
on the levels of the adjacent WZW models. In fact, each of the $\beta$-functions \eqref{e} are identical to the $\beta$-function of the model constructed in \cite{Georgiou:2017jfi}. 
Note also that the symmetry transformation \eqn{sysmmmr} induces the transformation
\be
\label{sysmmmr1}
 k^{(i)}\to k^{(n+3-i)}\, , \quad \l_0^{(i)}\to \big(\l_0^{(n+3-i)}\big)^{-1}\, ,\quad i=1,2,\dots , n\ . 
 \ee
 Therefore, the RG flow equations \eqn{e} are consistent with the transformation 
  of $\l_i$ in \eqn{sysmmmr}.

Clearly the CFT defined at the UV at $\l_i=0$ is the current algebra product $G_{k_1}\times G_{k_2}\times\dots\times G_{k_n}$. The RG flow is between this point and the IR fixed point. The IR CFT is reached when each of the deformation parameters $\l_i$ takes
a particular value, namely $\l_i=\l^{(i)}_0$ if $k_{i}>k_{i-1}$ or $\l_i=(\l_0^{(i)})^{-1}$ if $k_{i}<k_{i-1}$.
The details of the IR fixed point are, of course, sensitive to the ordering of the levels.
For example, for the case of $n=2$ it was shown in \cite{Georgiou:2017jfi} that for $k_1<k_2$ (which is an ordering we may always do with no loss of generality)
we obtain the following flow of CFTs from the UV point $(\l_1,\l_2)=(0,0)$ towards the IR $(\l_1,\l_2)=(\l_0,\l_0)$.
\begin{equation}
\label{f}
\boxed{
G_{k_1}\times G_{k_2}\ \Longrightarrow\ \frac{G_{k_1}\times G_{k_2-k_1}}{G_{k_2}}\times G_{k_2-k_1}}\ .
\end{equation}
The main goal of the present work is to be able to identify the nature of the CFTs in the IR regime. We will see that there is an increasing number of inequivalent IR CFTs as the number $n$ of coupled WZW models increases.
The nature of each of these CFTs crucially 
depends on the ordering of the levels or equivalently 
on the values that  each coupling  $\l_i$ takes at the IR fixed point. The reader might think that changing the order of the levels corresponds to the same RG flows, up to an appropriate renaming of the labels $i$. However,  this  is not the case since as we see from \eqref{d1}, only adjacent currents $J^i_{\pm}$ are coupled.
We will discover the very novel feature that for more than two coupled models the symmetries of the left and the right sectors of the IR CFTs   
 are not the same. Hence,  the IR CFT for $n=2$ depicted in \eqn{f} is a special case in which both sectors have the same symmetry. 
 Note that for $k_1=k_2$ the IR CFT in \eqn{f} seizes to have a meaning since the theory is driven to a strong
coupling regime in which it makes sense to consider the non-Abelian limit as in section 2 of  \cite{Georgiou:2017oly}.
Finally let us mention that \eqref{d} is a particular case of the most general multiparameter deformation of a $G_{k_1}\times\dots\times G_{k_n}$ CFT. The action was constructed in  \cite{Georgiou:2018gpe} and reads
\begin{equation}
\label{yes}
\begin{split}
S=&\sum_{i=1}^{n}S_{k_i}(g_i)+\sum_{i=1}^{n}\Big(\frac{k_i}{\pi}\int d^2\s\Tr(A^{(i)}_-\partial_+g_ig_i^{-1}-A^{(i)}_+g_i^{-1}\partial_-g_i+\\
&+A^{(i)}_-g_iA^{(i)}_+g_i^{-1})\Big)-\sum_{i,j=1}^{n}\frac{1}{\pi}\int d^2\s\sqrt{k_ik_j}\Tr(A^{(i)}_+\l_{ij}^{-1}A^{(j)}_-)\Big)\, ,
\end{split}
\end{equation}
where the inverses refer to the suppressed group
indices and not in the space of couplings with indices $i$ and  $ j$.
Choosing the entries of $\l^{-1}_{ij}$ as
\begin{equation}
\label{case}
\l_{n1}^{-1}=\l_1^{-1}\,, \qquad \l^{-1}_{ij}=\l_j^{-1}\delta_{i,j-1}
\end{equation}
 and redefining $A^{(n)}_+\to A^{(1)}_+$ and $A^{(i-1)}_+\to A^{(i)}_+$ this action becomes the one in \eqref{d}. 
Integrability of \eqn{yes} is ensured for a specific form of the matrix $\l^{-1}_{ij}$  taken to be isotropic in group space 
and given by \cite{Georgiou:2018gpe} 
\be
\label{lambda}
\begin{split}
& (\l_{ij})_{ab} = \d^{ab}\l_{ij} \ ,
\\
&\l^{-1}_{i1}\neq 0\ , \quad i=1,2,\dots ,n\!-\!1\ ,\qq
\l^{-1}_{nj}\neq 0 \ ,
\quad j=2,3,\dots ,n\ ,
\\
&\l^{-1}_{ij}= 0\,\,\,{\rm for\, all\, other\, entries}\ .
 \end{split}
\ee
The $\s$-model action \eqref{d}, is proven to be integrable in Appendix A  for diagonal matrices, 
i.e. $(\l_i)_{ab}=\l_i \d_{ab}$. Combining with \eqn{case} we conclude that the case at hand 
is a new integrable case compared to the one found in  \cite{Georgiou:2018gpe}.

\section{CFTs in the IR}
\label{CFTs at the IR regime}

In this section we will investigate the fixed point of the RG flow reached in the IR.

\subsection{The exact C-function and the central charge at the IR point}

A strong indication of the nature of the IR CFT may come from its central charge.  This is the value of Zamolodchikov C-function \cite{Zamolodchikov:1986gt} at the IR fixed point of the RG flow. The way to calculate C-functions as  non-trivial 
functions of the deformation parameters  in the  context of models obtained as deformations of  CFTs was put forward in \cite{Georgiou:2018vbb}, where the C-function calculated for the 
basic isotropic examples. Generalisations for generic deformation parameters can be found in \cite{Sagkrioti:2018abh}.

\no
The C-function was explicitly derived in \cite{Georgiou:2018vbb} for the $n=2$ case, so here we will just state the result as, the generalization for arbitrary $n$ is straightforward, the reason being that the deformation parameters $\l_i$ are decoupled. 
The C-function to $\mathcal{O}(1/k_i)$ in the large $k$-expansion reads \cite{Georgiou:2018vbb}
\begin{equation}\label{1}
C(\l_1,\dots,\l_n;\l_0^{(1)},\dots,\l_0^{(n)}) = n\dim G
- {c_G \dim G\ov 2 }\sum_{i=1}^{n}F(k_i,\l_i)\space\,,
\end{equation}
where  $c_G$ is the eigenvalue of the quadratic Casimir in the adjoint representation, i.e. $f_{acd}f_{bcd} = c_G \d_{ab}$ 
and
\begin{equation}\label{2}
F(k_i,\l_i)={1\ov k_i} +\l_i^3 {4 -\l_i(3-\l_i^2) (\l_0^{(i)}+(\l^{(i)}_0)^{-1})\ov k^{(i)}(1-\l_i^2)^3}\, ,
\quad i=1,2,\dots , n \ .
\end{equation}
When \eqref{1} is evaluated  at a fixed point it will give the large $k^{(i)}$ expansion of the central charge of the CFT defined at this point. For example, at the UV point where all the couplings are zero, we get
\begin{equation}
c_{\rm{UV}}=n\dim G-\frac{c_G\dim G}{2}\sum_{i=1}^{n}\frac{1}{k_i}\,,
\end{equation}
which is indeed the large $k_i$ expansion of the central charge of a $G_{k_1}\times\dots\times G_{k_n}$ CFT.

Before we proceed to compute the central charge of the CFTs  at the IR regime for the cases $n=2,3,4$, which are the cases we will present in full detail in the rest of the paper, we would like to make a few remarks.
The parameters $\l_i$ run independently under the RG flow \eqref{e} and the UV fixed point is reached when all deformation parameters equal zero. In the IR fixed point each $\l_i$ reaches the smaller of $\l_0^{(i)}$ or its inverse, $(\l_0^{(i)})^{-1}$, depending on the ordering of the levels.
Due to the specific form of $\l_0^{(i)}$ it turns out that different orderings of the levels $k_i$ may correspond to the same IR 
fixed point. For example, for the case of $n=4$ the orderings $k_1<k_3<k_2<k_4$ and $k_1<k_3<k_4<k_2$ correspond to the same IR fixed point and in fact to the same RG flow\footnote{It is easy to see this at the linearized level. Indeed, note that the operator driving the CFT away from the UV point, up to first order in $\l_i$, from \eqn{d1}  is 
\begin{equation}
\label{3.4}
L_{\rm{int}}=k^{(2)} \l_2\Tr(J^{(2)}_{+}J^{(1)}_{-})+k^{(3)}\l_3\Tr(J^{(3)}_{+}J^{(2)}_{-})+k^{(4)}\l_4\Tr(J^{(4)}_{4+}J^{(3)}_{-})
+k^{(1)}\l_1 \Tr(J^{(1)}_{+}J^{(4)}_{-})\,.
\end{equation}
Under the transformation \eqn{sysmmmr} for $n=4$ we have that $\s_+\leftrightarrow\s_-$,
$g_2\leftrightarrow g_4^{-1}$, $k_2\leftrightarrow k_4$, $\l_1\leftrightarrow \l_2$ and $\l_3\leftrightarrow \l_4$,
then \ref{3.4} remains invariant.}
\begin{equation}
\label{3.6}
{\rm UV}: (0,0,0,0)\quad \Longrightarrow\quad  {\rm IR}: ((\l^{(1)}_0)^{-1},\l^{(2)}_0,(\l^{(3)}_0)^{-1},\l^{(4)}_0)\, .
\end{equation}\par
\noindent 
Thus we classify every CFT at the IR in terms of the different combinations of the values of the deformation parameters (that is whether $\l_i$ is equal to $\l_0^{(i)}$ or $(\l_0^{(i)})^{-1}$) and not in terms of the orderings of the levels. We assume without generality loss, that $k_1$ is the smallest level among them all. 
Then, in the IR fixed point we necessarily have that $\l_1=(\l_0^{(1)})^{-1}=\sqrt{\frac{k_1}{k_n}}$ and that $\l_2=\l_0^{(2)}=\sqrt{\frac{k_1}{k_2}}$.
For a given $n$ we have $(n-1)!$ different orderings of the levels $k_i$, $i=2,3,\dots ,n$ which correspond to $2^{n-2}$ different IR fixed points, thus CFTs. This is the case because for a general number $n$ of WZW models 
\eqref{3.6} will have $n-2$ independent entries to each of which one can assign a $\l_0$ or a $\l_0^{-1}$.
In the following sections we will see that the independent IR CFTs are further
reduced in number, due to interrelations using the parity transformation \eqn{sysmmmr}, \eqn{sysmmmr1}.

We will organize the notation of the IR fixed points by assigning to every $\l_0^{(i)}$ an arrow pointing upwards and to its inverse an arrow pointing downwards. We disregard the ones with $i=1\,,2$, since their direction is fixed for every ordering, the first one is always pointing downwards and the second one always upwards. 
For example, in the $n=3$ case we have two possible CFTs defined at the IR fixed points $((\l_0^{(1)})^{-1},\l_0^{(2)},\l_0^{(3)})=(\uparrow)$ and $((\l_0^{(1)})^{-1},\l_0^{(2)},(\l_0^{(3)})^{-1})=(\downarrow)$. For $n=4$ there are
six  different orderings of the levels which correspond to four different IR fixed points
\begin{equation}
\label{upd4}
\begin{split}
&(\uparrow,\uparrow)= ((\l_0^{(1)})^{-1},\l^{(2)}_0,\l^{(3)}_0,\l^{(4)}_0)\,,\\
&(\downarrow,\uparrow)= ((\l_0^{(1)})^{-1},\l^{(2)}_0,(\l^{(3)}_0)^{-1},\l^{(4)}_0)\,,\\
&(\uparrow,\downarrow)= ((\l_0^{(1)})^{-1},\l^{(2)}_0,\l^{(3)}_0,(\l^{(4)}_0)^{-1})\,,\\
&(\downarrow,\downarrow)= ((\l_0^{(1)})^{-1},\l^{(2)}_0,(\l^{(3)}_0)^{-1},(\l^{(4)}_0)^{-1})\,.
\end{split}
\end{equation}
In our notation for $n=2$, the IR fixed point $(\l_1,\l_2)=(\l_0,\l_0)$ will be denoted by $(0)$. 

\no
In order to recognize the IR fixed point CFT we will utilize the information contained in the large level 
expansion of the $C$-function \eqref{1} when this is evaluated at the end of the flow in the IR. 
For the cases with $n=2,3,4$ we present the result in table \ref{table:e'}. 

\begin{table}[h]
\renewcommand{\arraystretch}{2.5}
	\centering
	\begin{tabular}{ |c||c|c| }
		\hline
		n & IR & $\sum_{i=1}^{n}F(k_i,\l_i)\big |_{\rm IR}$ \\
		\hline
		\multirow{1}{*}{$2$} & $(0)$& $\displaystyle{\frac{1}{k_1}+\frac{1}{k_2-k_1}-\frac{1}{k_2}}+\frac{1}{k_2-k_1}$\\
		\arrayrulecolor{black}\hline \arrayrulecolor{black}
		\multirow{2}{*}{$3$} & $(\uparrow)$ & $\displaystyle{\frac{1}{k_1}+\frac{1}{k_3-k_1}-\frac{1}{k_3}+\frac{1}{k_2-k_1}+\frac{1}{k_3-k_2}}$\\\cline{2-3}
		& $(\downarrow)$  & $\displaystyle{\frac{1}{k_1}+\frac{1}{k_2-k_1}-\frac{1}{k_2}+\frac{1}{k_3-k_1}+\frac{1}{k_2-k_3}}$\\ \arrayrulecolor{black}\hline \arrayrulecolor{black}
		\multirow{4}{*}{$4$} & $(\uparrow,\uparrow)$ & $\displaystyle{\frac{1}{k_1}+\frac{1}{k_4-k_1}-\frac{1}{k_4}+\frac{1}{k_2-k_1}+\frac{1}{k_3-k_2}+\frac{1}{k_4-k_3}}$\\\cline{2-3}
		& $(\downarrow,\uparrow)$ & $\displaystyle{\frac{1}{k_1}+\frac{1}{k_4-k_1}-\frac{1}{k_4}+\frac{1}{k_3}+\frac{1}{k_2-k_3}-\frac{1}{k_2}+\frac{1}{k_2-k_1}+\frac{1}{k_4-k_3}}$\\\cline{2-3}
		& $(\uparrow,\downarrow)$ & $\displaystyle{\frac{1}{k_1}+\frac{1}{k_4-k_1}+\frac{1}{k_3-k_4}-\frac{1}{k_3}+\frac{1}{k_2-k_1}+\frac{1}{k_3-k_2}}$\\\cline{2-3}
		& $(\downarrow,\downarrow)$ &$\displaystyle{\frac{1}{k_1}+\frac{1}{k_2-k_1}-\frac{1}{k_2}+\frac{1}{k_4-k_1}+\frac{1}{k_3-k_4}+\frac{1}{k_2-k_3}}$\\
		\hline
	\end{tabular}
	\caption{Large $k_i$ expansion of the central charge of the IR CFTs for $n=2,3,4$.}
	\label{table:e'}
\end{table}
\no
We will assume that the CFTs we are after, are of the current algebra type or of the coset type, an assumption that will be 
further supported later in the paper.
Under this assumption, we may assign the various $1/k$-factors according to the rule
\be
\begin{split}
& \frac{1}{k_i}\!-\!{\rm term} \quad \Longrightarrow\quad  G_{k_i}\!-\!{\rm chiral\ current\ symmetry}\ ,
\\
&
\bigg(\frac{1}{k_i}+\frac{1}{k_{j}}-\frac{1}{k_i+k_j}\bigg)\!-\!{\rm term}  \quad \Longrightarrow\quad \frac{G_{k_i}\times G_{k_{j}}}{G_{k_{i}+k_j}}\!-\!{\rm chiral\ coset\ symmetry}\ .
\end{split}
\ee
Then, we may easily  deduce  that for $n=2$ the symmetry of the CFT at $(0)$ is that in  \eqref{f}, as shown in \cite{Georgiou:2017jfi}. However, this reasoning alone does not uniquely fix the symmetry of the IR CFT for higher values of $n$. 

\noindent For example, the CFT, even at the IR fixed point $(\uparrow)$, can be equally well described with both types of conformal symmetries 
\begin{equation}
\frac{G_{k_1}\times G_{k_3-k_1}}{G_{k_3}}\times G_{k_2-k_1}\times G_{k_3-k_2}\quad\text{or}\quad
\frac{G_{k_1}\times G_{k_2-k_1}\times G_{k_3-k_2}}{G_{k_3}}\times G_{k_3-k_1}\,.
\end{equation}
These CFTs are clearly different but nevertheless have the same central charge.  
 We will see in the following, that  both  contribute to the  underlying conformal symmetry of the IR  theory, one to the left and the other to the right sector.
 The next sections will be dedicated to the explicit derivation of the conformal symmetry of the IR CFTs presented in table \eqref{table:e'}.

\subsection{Identifying the left-right current chiral symmetries of the IR CFTs}

As a warm up  we start with the easiest case having two deformation parameters. 
The chiral symmetries of the CFT defined at the IR point $(0)$, are clearly exhibited if we set $(\l_1,\l_2)=(\l_0,\l_0)$ in \eqref{d}. In this case the action takes the form (after relabeling the gauge fields  $A_\pm^{(1)}$ and $A_\pm^{(2)}$ by 
$A_\pm $ and $B_\pm$, respectively)  
\begin{equation}\label{g}
\begin{split}
&S=S_{k_1}(g_1)+S_{k_2}(g_2)+{k_1\ov \pi}\int d^2\s\, \Tr(A_-\partial_+g_1g_1^{-1}-B_+g_1^{-1}\partial_-g_1+A_-g_1B_+g_1^{-1})\\
&\quad +{k_2\ov \pi}\int d^2\s\Tr(B_-\partial_+g_2g_2^{-1}-A_+g^{-1}_2\partial_-g_2+B_-g_2A_+g_2^{-1}-A_+A_--B_+B_-)\, .
\end{split}
\end{equation}
We next parametrize $A_{\pm}$ and $B_{\pm}$ in terms of $a_{\pm}$ and $b_{\pm}$, taking values in $G$, as 
\begin{equation}
\label{i}
A_{\pm}=\partial_{\pm}a_{\pm}a_{{\pm}}^{-1}\, ,\quad B_{\pm}=\partial_{\pm}b_{\pm}b^{-1}_{\pm}\,.
\end{equation}
Then,
using the Polyakov-Weigmann (PW) identity we can rewrite \eqref{g} as\footnote{In our conventions the PW identity is 
\begin{equation}
\label{PWW}
S_k(g_2g_1)=S_k(g_1)+S_k(g_2)+\frac{k}{\pi}\int d^2\s\, \Tr(J_{1+}J_{2-})\,.
\end{equation}}
\begin{equation}
\label{h}
\begin{split}
& S=S_{k_1}(a^{-1}_-g_1b_+)+S_{k_2}(b^{-1}_-g_2a_+)-S_{k_2}(a_-^{-1}a_+)-S_{k_2}(b_-^{-1}b_+)
\\
& 
\qq +S_{k_2-k_1}(a_-^{-1})+S_{k_2-k_1}(b_+)\, .
\end{split}
\end{equation}
Since we can rewrite \eqref{g} as a sum of WZW actions of independent fields the conformal invariance of the model is preserved at the quantum level.
\par
\noindent
In order to clarify the type of chiral symmetry of this CFT we make two observations:
\textbullet \,\,\,The first one is that under the transformation
\begin{equation}
\begin{split}
&(g_1, g_2)\to (h^{-1}_1g_1h_2, h^{-1}_2g_2h_1)\, ,
\\
&(a_{\pm}, b_{\pm})\to (h^{-1}_1a_{\pm},h^{-1}_2b_{\pm})\,,
\end{split}
\end{equation}
with $h_1, h_2\in G$, the infinitesimal variation of the action is
\begin{equation}\label{j}
\delta S=\frac{k_2-k_1}{\pi}\int d^2\, \s\Tr(A_-\partial_+\e_1+B_+\partial_-\e_2)\,,
\end{equation}
where $\e_1,\e_2\in \mathcal{L}(G)$. This result  indicates that the action \eqref{h} possesses two chiral symmetries, i.e $\e_1=\e_1(\s_-), \e_2=\e_2(\s_+)$, with associated  chiral and antichiral currents $B_+$ and $A_-$ at level $(k_2-k_1)$.
In further support of this, the equations of motion of the fields $A_-$, $B_+$, as it is shown in \eqref{14} of Appendix A, in the case of arbitrary $\l_1, \l_2$ are given by 
\begin{equation}
\partial_+A_-=\frac{1-\l^{-1}_0\l_1}{1-\l^2_1}\big[A_+,A_-\big]\, , \quad \partial_-B_+=-\frac{1-\l^{-1}_0\l_2}{1-\l^2_2}\big[B_+,B_-\big]\,.
\end{equation}
Hence, at the IR fixed point $(\l_1,\l_2)=(\l_0,\l_0)$ the fields indeed become chiral and antichiral.

\noindent 
\textbullet\,\,\,The second observation is that , as it is shown in Appendix B, in the IR fixed point $(0)$ the equal time Poisson brackets of the two fields $A_-$, $B_+$ reduce to two independents Kac--Moody algebras with central extension $(k_2-k_1)$
(copied for convenience from \eqn{B12} and \eqn{B13})
\begin{equation}
\label{k}
\begin{split}
&\Big\{A^a_-(\s),A^b_-(\s')\Big\}=\frac{i}{k_2-k_1}f^{abc}A^c_-\delta_{\s\s'}+\frac{1}{k_2-k_1}\delta^{ab}\delta'_{\s\s'}\, ,
\\
&\Big\{B^a_+(\s),B^b_+(\s')\Big\}=\frac{i}{k_2-k_1}f^{abc}B^c_+\delta_{\s\s'}-\frac{1}{k_2-k_1}\delta^{ab}\delta'_{\s\s'}\,.
\end{split}
\end{equation}

\no
From the above we deduce that  the chiral symmetry of the action \eqref{g} is
\begin{equation}\label{l}
\boxed{\big(G_{k_2-k_1}\big)_L\times\big(G_{k_2-k_1}\big)_R}\,.
\end{equation}
\subsubsection{Current chiral symmetries of the IR CFTs for arbitrary $n$}
This analysis is easily generalized to arbitrary $n$. Suppose the ordering of the levels is $k_1<k_2<\dots<k_n$. Then the  IR conformal point is reached for 
\begin{equation}
(\l_1,\l_2,\dots,\l_n)=((\l^{(1)}_0)^{-1},\l^{(2)}_0,\dots,\l^{(n)}_0)\,.
\end{equation} 
Furthermore, the action \eqref{d} becomes
\begin{equation}\label{m}
\begin{split}
S&=\sum_{i=1}^{n}S_{k_i}((a^{(i)}_-)^{-1}g_ia^{(i+1)}_+)-S_{k_n}((a^{(1)}_-)^{-1}a^{(1)}_+)-\sum_{i=2}^{n}S_{k_i}((a^{(i)}_-)^{-1}a^{(i)}_+)+\\
&\quad +S_{k_n-k_1}((a^{(1)}_-)^{-1})+\sum_{i=2}^{n}S_{k_i-k_{i-1}}(a^{(i)}_+)\,,
\end{split}
\end{equation}
where we used again the Polyakov--Weigman identity \eqn{PWW} and we have parametrized each of the fields $A^{(i)}_{\pm}\in\mathcal{L}(G)$ in terms of $a^{(i)}_{\pm}\in G$ similarly to  \eqref{i}. As before  under the transformation $g_i\to h^{-1}_ig_ih_{i+1}$, $a^{(i)}_{\pm}\to h^{-1}_ia^{(i)}_{\pm}$,  the infinitesimal variation of \eqref{m} reads
\begin{equation}\label{n}
\delta S={k_n-k_1\ov \pi} \int d^2\s\, \Tr A^{(1)}_-\partial_+\e_1+\sum_{i=2}^{n}{k_i-k_{i-1}\ov\pi} \int d^2\s\, \Tr A^{(i)}_+\partial_-\e_i\,.
\end{equation}
It is a lengthy, albeit straightforward calculation in the canonical formalism, to show that the fields $A_+^{(1)}$ and
 $A^{(i)}_+$ with $i=2,3,\dots,n$ obey $n$ independent Kac--Moody algebras with central extensions $k_n-k_1$ and $k_i-k_{i-1}$, respectively. Thus the chiral symmetry of \eqref{m} is based on the following products
\begin{equation}\label{o}
\boxed{\big(G_{k_2-k_1}\times\dots\times G_{k_n-k_{n-1}}\big)_{L}\times\big(G_{k_n-k_1}\big)_{R}}\, .
\end{equation}

If we now make a small modification in the ordering of the levels, for example if we pick $k_i<k_{i-1}$ the IR conformal point is reached for
\begin{equation}\label{nn'}
(\l_1,\l_2,\dots,\l_i,\dots,\l_n)=((\l^{(1)}_0)^{-1},\l^{(2)}_0,\dots,(\l^{(i)}_0)^{-1},\dots,\l^{(n)}_0)\,,
\end{equation}
Under the same transformation as before the infinitesimal variation of the action at the IR point \eqref{nn'} is
\begin{equation}\label{n'}
\begin{split}
\delta S&={k_n-k_1\ov \pi}\int d^2\s\, \Tr A^{(1)}_-\partial_+\e_1+{k_{i-1}-k_i\ov \pi}\int d^2\s\, \Tr A^{(i)}_-\partial_+\e_i+\\
&+\sum_{j=2,\atop j\ne i}^{n}(k_j-k_{j-1})\int d^2\s\, \Tr A^{(i)}_+\partial_-\e_i\, .
\end{split}
\end{equation}
Similar reasoning as before indicates that the chiral symmetry of the CFT is
\begin{equation}\label{o'}
\boxed{\big(G_{k_2-k_1}\times\dots\times\hat{G}_{k_{i}-k_{i-1}}\times\dots\times G_{k_n-k_{n-1}}\big)_{L}\times \big(G_{k_n-k_1}\times G_{k_{i-1}-k_{i}}\big)_{R}}\,,
\end{equation}
where the hat above $G_{k_{i}-k_{i-1}} $ indicates that this term is missing from the sequence. In fact it appears 
in the right sector with the opposite sign for the level which is now positive.

It is now straightforward to identify the chiral symmetries of the conformal field theories at the fixed points for $n=2,3,4$, which we present in table \ref{table:eee'}. 
\begin{table}[h]
\renewcommand{\arraystretch}{1.4}
	\centering
		\begin{tabular}{ |p{1.5cm}||p{7.5cm}|}
			\hline
			IR point     & Chiral symmetries\\
				\hline
			$\displaystyle{(0)}$   &$\displaystyle{\big(G_{k_2-k_1}\big)_{L}\times\big(G_{k_2-k_1}\big)_{R} }$\\
			\hline
			$\displaystyle{(\uparrow)}$ &$\displaystyle{\big(G_{k_2-k_1}\times G_{k_3-k_2}\big)_{L}\times\big(G_{k_3-k_1}\big)_{R}}$ \\
			\hline
			$\displaystyle{(\downarrow)}$ &$\displaystyle{\big(G_{k_2-k_1}\big)_{L}\times\big(G_{k_3-k_1}\times G_{k_2-k_3}\big)_{R}}$ \\
			\hline
			$\displaystyle{(\uparrow,\uparrow)}$   &$\displaystyle{\big(G_{k_2-k_1}\times G_{k_3-k_2}\times G_{k_4-k_3}\big)_L\times\big(G_{k_4-k_1}\big)_{R}}$  \\
			\hline
			$\displaystyle{(\downarrow,\uparrow)}$ &$\displaystyle{\big( G_{k_2-k_1}\times G_{k_4-k_3}\big)_{L}\times\big(G_{k_4-k_1}\times G_{k_2-k_3}\big)_{R}}$ \\
			\hline
			$\displaystyle{(\uparrow,\downarrow)}$ &$\displaystyle{\big(G_{k_2-k_1}\times G_{k_3-k_2}\big)_{L}\times\big(G_{k_4-k_1}\times G_{k_3-k_4}\big)_{R}}$\\
			\hline
			$\displaystyle{(\downarrow,\downarrow)}$ &$\displaystyle{ \big(G_{k_2-k_1}\big)_{L}\times \big(G_{k_2-k_3}\times G_{k_3-k_4}\times G_{k_4-k_1}\big)_{R}}$\\
			\hline
		\end{tabular}
	\caption{Chiral symmetries of the CFTs defined at the IR points for $n=2,3,4$.}
	\label{table:eee'}
\end{table}
The above symmetries are in agreement, as far as the chiral symmetries are concerned, with the central charges presented in table \ref{table:e'}. 
Notice also that, the case of $n=2$ is a special case, since the left and right chiral algebras coincide. 
This will become clearer in the following.

\par
\subsection{The full conformal symmetry}
In order to set the grounds for our arguments we will develop our formalism for the case of $n=2$ and then we will extend it to larger values of $n$.
Our strategy will be to determine an action which after a gauge fixing will give the action of the IR CFT obtained after integrating out a subset of the gauge fields $A_\pm^{(i)}$.
The symmetries of the IR CFT can then be easily read from the former action.
\subsubsection{The conformal symmetry for $n=2$}
The starting point will be the action \eqref{g} or equivalently \eqref{h}. Since the action is quadratic in the fields  we can perform the integration over them. Integrating out $B_{\pm}$ we find (this produces a trivial determinant, i.e. a trivial dilaton field)
\begin{align}\label{p}
S=S_{k_1}(g,A)+S_{k_2-k_1}(g_2)-{k_2-k_1\ov \pi}\int d^2\s\, \Tr(A_+g_2^{-1}\partial_-g_2-A_+A_-)\,,
\end{align}
where $S_{k_1}(g,A)$ stands for the standard diagonal vector gauged WZW model and we used the invariance of the 
path integral measure to replace the integral over the group variables $g_1$ and $g_2$ by one over $g=g_1g_2$ and $g_2$. Using the parametrization \eqref{i} and the PW formula we find that \eqref{p} can be rewritten as
\begin{equation}\label{q}
S=S_{k_1}(a^{-1}_-ga_+)+S_{k_2-k_1}(g_2a_+)+S_{k_2-k_1}(a^{-1}_-)-S_{k_2}(a^{-1}_-a_+)\,.
\end{equation}
We would like to identify \eqref{q} as the gauge fixed version of an action which realizes the chiral symmetries and central charge presented in the previous sections for the case of the IR point $(0)$. Suppose that we define the following action
\begin{equation}\label{r}
S=S_{k_1}(g_1)+S_{k_2-k_1}(g_2)+S_{k_2-k_1}(g_3)\, ,
\end{equation}
which is the WZW model defined on the product group $G=G_{k_1}\times G_{k_2-k_1}\times G_{k_2-k_1}$. This model clearly has a global $G_L\times G_R$ symmetry. We now gauge the anomaly free subgroup $H\subset G_L\times G_R$, defined as
\begin{equation}\label{s}
H: (g_1,g_2,g_3)\to (h^{-1}g_1h,g_2h,h^{-1}g_3)\,,
\end{equation}
with $h$ some constant element in $G$. The gauged action then takes the form
\begin{equation}\label{t}
S_{G/H}=S_{k_1}(a^{-1}_-g_1a_+)+S_{k_2-k_1}(g_2a_+)+S_{k_2-k_1}(a^{-1}_-g_3)-S_{k_2}(a^{-1}_-a_+)\,,
\end{equation}
where in order to find it  we have introduced the gauge fields $A_{\pm}=\partial_{\pm}a_{\pm}a_{{\pm}}^{-1}$
with $a_\pm \in G$.
The gauge  symmetry of \eqref{t} is given by \eqn{s}, but with a local element $h(\s_+,\s_-)\in G$ and the additional
transformation  $a_{\pm}\to h^{-1}a_{\pm}$.
In addition, there are the chiral symmetries given by
\begin{equation}
\label{u}
(g_2,g_3)\to \big(\Omega^{-1}_L(\s_+)g_2,g_3\Omega_R(\s_-)\big)\, , \quad \Omega_L(\s_+), \Omega_R(\s_-)\in G\,.
\end{equation}
From the above we deduce that conformal symmetry of \eqref{t} is
\begin{equation}
\label{v}
\left.\frac{G_{k_1}\times G_{k_2-k_1}}{G_{k_2}}\times G_{k_2-k_1}\right|_{L}\ ,\qquad 
\left.\frac{G_{k_1}\times G_{k_2-k_1}}{G_{k_2}}\times G_{k_2-k_1}\right|_{R}\,.
\end{equation}
The two coset type symmetries in the left and right sector appearing in \eqref{v} correspond to the transformation in  \eqref{s} while the remaining two chiral symmetries at level $k_2-k_1$ to \eqref{u}. This is in agreement with the previous section in which we saw that the two chiral symmetries are generated by the fields $B_+, A_-$ which satisfy two independent Kac-Moody algebras at level $k_2-k_1$. Since the action \eqref{t} is gauge invariant we may fix the gauge. Choosing to fix the gauge by setting $g_3=\mathbb{1}$, the action \eqref{t} becomes precisely \eqref{q}. Choosing instead to fix the gauge by setting $g_2=\mathbb{1}$, the  resulting action would be the same as the one obtained from \eqref{g} but with the $A_{\pm}$'s  integrated out instead of the $B_{\pm}$.
Note that, we could not choose $g_1=\mathbb{1}$ as the gauge group acts vectorially on $g_1$.

Thus, since the procedure of integrating out does not affect the conformal symmetry of a theory, we conclude that the conformal symmetry of the CFT defined at the IR point $(0)$  and which is
described by the action\eqref{g} 
is the one given in \eqref{v} or equivalently in \eqn{f}.

\subsubsection{The conformal symmetry for $n=3$}
In this case we have three levels  $k_1, k_2, k_3$. Since $k_1$ is the smallest one fixed, we have two different IR fixed points, $(\uparrow)$ and $ (\downarrow)$ which correspond to $k_2 < k_3$ and $k_3 < k_2$, respectively.

\noindent
$\bullet$
The CFT defined at $(\uparrow)$ is found from \eqn{d} evaluated at the IR fixed point at which 
$(\l_1,\l_2,\l_3)=((\l_0^{(1)})^{-1},\l_0^{(2)},\l_0^{(3)})$.
After a relabeling of the gauge fields $A_\pm^{(1)}, A_\pm^{(2)}, A_\pm^{(3)}$ by $A_\pm,B_\pm, C_\pm$ 
and the use of the PW identity it is described by the action
\begin{equation}\label{w}
\begin{split}
S=&S_{k_1}(a^{-1}_-g_1b_+)+S_{k_2}(b^{-1}_-g_2c_+)+S_{k_3}(c^{-1}_-g_3a_+)+S_{k_3-k_1}(a^{-1}_-)+S_{k_2-k_1}(b_+)\\
&+S_{k_3-k_2}(c_+)-S_{k_2}(b^{-1}_-b_+)-S_{k_3}({c^{-1}_-c_+})-S_{k_3}(a^{-1}_-a_+)\,,
\end{split}
\end{equation}
where we have used \eqn{i} and the group elements $c_{\pm}$ correspond to the third gauge field $C_{\pm}$. 
Performing the integration over the fields $(b_{\pm},c_{\pm})$ (equivalently $(B_\pm , C_\pm)$) we find
\be
\label{x}
\begin{split}
S&=S_{k_1}(f)+S_{k_2-k_1}(g)+S_{k_3-k_2}(g_3)+\frac{k_1}{\pi}\int d^2\s\, \Tr\big(A_-\partial_+ff^{-1}
-A_+f^{-1}\partial_-f
\\
& \quad +A_-fA_+f^{-1}\big) -\frac{k_2-k_1}{\pi}\int d^2\s\, \Tr\big(A_+g^{-1}\partial_-g\big)
\\
&\quad -\frac{k_3-k_2}{\pi}\int d^2\s\, \Tr\big(A_+g_3^{-1}\partial_-g_3\big) -\frac{k_3}{\pi}\int d^2\s\, \Tr(A_+A_-)
\\
&=
S_{k_1}(a^{-1}_-fa_+)+S_{k_2-k_1}(ga_+)+S_{k_3-k_2}(g_3a_+)-S_{k_3}(a^{-1}_-a_+)+S_{k_3-k_1}(a^{-1}_-)\,,
\end{split}
\ee
where we replaced  the group elements $g_1$ and $g_2$  with the group elements $f=g_1g_2g_3$, $g=g_2g_3$. Using a similar reasoning as before for the $n=2$ case, we find that \eqref{x} is the gauge fixed version, with  $g_4=\mathbb{1}$,
 of the gauged WZW action
 \be
 \label{y}
\begin{split}
S_{G/H}&=S_{k_1}(g_1)+S_{k_2-k_1}(g_2)+S_{k_3-k_2}(g_3)+S_{k_3-k_1}(g_4)\\
&\quad +\frac{k_1}{\pi}\int d^2\s\, \Tr\big(A_-\partial_+g_1g_1^{-1}-A_+g_1^{-1}\partial_-g_1+A_-g_1A_+g_1^{-1}\big)
\\
&\quad -\frac{k_2-k_1}{\pi}\int d^2\s\, \Tr\big(A_+g_2^{-1}\partial_-g_2\big)
-\frac{k_3-k_2}{\pi}\int d^2\s\, \Tr\big(A_+g_3^{-1}\partial_-g_3\big)
\\
&\quad +\frac{k_3-k_1}{\pi}\int d^2\s\, \Tr\big(A_-\partial_+g_4g_4^{-1}\big)
-\frac{k_3}{\pi}\int d^2\s\, \Tr\big(A_+A_-\big)
\\
& =S_{k_1}(a^{-1}_-g_1a_+)+S_{k_2-k_1}(g_2a_+)+S_{k_3-k_2}(g_3a_+)
\\
& \qq\quad +S_{k_3-k_1}(a^{-1}_-g_4)-S_{k_3}(a^{-1}_-a_+)\,,
\end{split}
\ee
where $H$ acts as
\begin{equation}
\label{g1234}
H: (g_1,g_2,g_3,g_4)\to (h^{-1} g_1h,g_2h,g_3h,h^{-1}g_4)\, ,\quad h\in G\,.
\end{equation}
Notice again that we can fix the gauge in \eqref{y}
by choosing $g_3=\mathbb{1}$. If we do so, the resulting action can be obtained from \eqref{w}, if we integrate out the fields $(a_{\pm},b_{\pm})$. Indeed, if we perform such an integration in \eqref{w} we find again \eqn{x} with some field
relabeling.  

\no
Turning our attention to the symmetries of \eqref{y}, the gauge symmetry is given by \eqn{g1234} with a local element
$h(\s_+,\s_-)\in G$ and in addition with $a_{\pm}\to h^{-1}a_{\pm}$.
The chiral symmetries are
\begin{equation}
\label{ccs}
(g_2,g_3,g_4)\to \big(\Omega^{-1}_L(\s_+)g_2,\tilde{\Omega}^{-1}_L(\s_+)g_3,g_4\Omega_R(\s_-)\big)\,.
\end{equation}
Thus we conclude that the conformal symmetry of the CFT defined at the IR fixed point $(\uparrow)$ is given by
\begin{equation}
\label{z}
\left.\frac{G_{k_1}\times G_{k_3-k_1}}{G_{k_3}}\times G_{k_2-k_1}\times G_{k_3-k_2}\right|_{L}\ ,\qq
\left. \frac{G_{k_1}\times G_{k_2-k_1}\times G_{k_3-k_2}}{G_{k_3}}\times G_{k_3-k_1}\right|_{R}\,.
\end{equation}
Notice the asymmetry between the left and right sectors. Despite this asymmetry the left and right moving Virassoro algebra possess the same central charge, which agrees with the central charge presented in section 2.1.
The result \eqref{z} is also in agreement with the chiral symmetries that we derived previously.

\no
The attentive reader might wonder what would happen if in \eqn{w} we integrate out the fields $(a_{\pm}, c_{\pm})$
(equivalently $(A_\pm,C_\pm)$) which is the only other possibility, presumably and in principle equivalent to the ones 
considered leading to the same conclusion. 
We found terms assigned with ratios of levels, i.e. $\displaystyle{{k_1k_2\ov k_3} 
g_2^{-1}\partial_-g_2D_3\partial_+g_1g_1^{-1}}$. Such terms cannot be factorized using the PW identity so the resulting action can not be easily associated with a CFT.

\no
$\bullet$
We now turn to the other fixed point, denoted by
$(\downarrow)$. The action of the corresponding CFT is given by
\begin{equation}
\begin{split}
S=&\mathcal{S}_{k_1}(a^{-1}_-g_1b_+)+S_{k_2}(b^{-1}_-g_2c_+)+S_{k_3}(c^{-1}_-g_3a_+)+S_{k_3-k_1}(a^{-1}_-)+S_{k_2-k_1}(b_+)\\
&+S_{k_2-k_3}(c^{-1}_-)-S_{k_2}(b^{-1}_-b_+)-S_{k_2}({c^{-1}_-c_+})-S_{k_3}(a^{-1}_-a_+)\, ,
\end{split}
\end{equation}
after substituting  in \eqref{d} the values $(\l_1,\l_2,\l_3)=((\l_0^{(1)})^{-1},\l_0^{(2)},(\l_0^{(3)})^{-1})$.
The procedure is identical to that before so that we will skip most of the details. 
If we integrate out either $(a_{\pm}, b_{\pm})$ or $(a_{\pm}, c_{\pm})$  and identify the resulting action as a gauge fixed version of a gauge invariant one, depending on an additional group element, we conclude that the conformal symmetry of the CFT defined at the IR point $(\downarrow)$ is given by
\begin{equation}
\label{aa}
\left.\frac{G_{k_1}\times G_{k_3-k_1}\times G_{k_2-k_3}}{G_{k_2}}\times G_{k_2-k_1}\right|_{L}\ ,
\qq
\left. \frac{G_{k_1}\times G_{k_2-k_1}}{G_{k_2}}\times G_{k_2-k_3}\times G_{k_3-k_1}\right|_{R}\,.
\end{equation}
Again the left and right moving Virassoro algebra possess the same central charges. It is needless to say that result \eqref{aa} agrees with the central charge and the chiral symmetries presented in tables \ref{table:e'} and \ref{table:eee'}
For the convenience of the reader the IR CFTs for $n=3$ are listed in the table below 
\begin{table}[h!]
	\begin{footnotesize}
		\renewcommand{\arraystretch}{2.7}
		\begin{center}
			\begin{tabular}{ |p{0.5cm}||p{5.2cm}|p{5.2cm}| }
				\hline
				IR     & Left sector & Right sector\\
				\hline
				$\displaystyle{(\uparrow)}$   &$\displaystyle{\frac{G_{k_1}\times G_{k_3-k_1}}{G_{k_3}}\times G_{k_2-k_1}\times G_{k_3-k_2}}$   &$\displaystyle{\frac{G_{k_1}\times G_{k_2-k_1}\times G_{k_3-k_2}}{G_{k_3}}\times G_{k_3-k_1}}$\\
				\hline
				$\displaystyle{(\downarrow)}$ &$\displaystyle{\frac{G_{k_1}\times G_{k_3-k_1}\times G_{k_2-k_3}}{G_{k_2}}\times G_{k_2-k_1}}$  &$\displaystyle{\frac{G_{k_1}\times G_{k_2-k_1}}{G_{k_2}}\times G_{k_3-k_1}}\times G_{k_2-k_3}$ \\
				\hline
			\end{tabular}
		\end{center}
		\caption{The conformal field theories at the IR fixed points for $n=3$.}
		\label{table:eeee'}
	\end{footnotesize}
\end{table}

\no
Note that, the two IR CFTs we found for $n=3$ are not independent. They are instead, related under the renaming $k_2\leftrightarrow k_3$ and the parity transformation in order interchange the left with the right sector.

\subsubsection{The conformal symmetry for $n=4$}
For $n=4$ we have four different IR fixed points. They correspond to different CFTs and in table \ref{table:ee'}, we present the 
associated symmetries for each one of them. Needless to say that the central charges agree,
in the large level limit, with the ones calculated in section 2.1.
We will perform a detailed derivation of the conformal symmetry of the field theories defined at $(\downarrow,\uparrow)$ and $ (\uparrow,\downarrow)$ with the other two being straightforward generalizations of the $n=3$ cases.

\begin{table}[h]
	\begin{footnotesize}
		\renewcommand{\arraystretch}{2.7}
		\begin{center}
			\begin{tabular}{ |p{0.8cm}||p{7.2cm}|p{7.2cm}| }
				\hline
				IR     & Left sector & Right sector\\
				\hline
				$\displaystyle{(\uparrow,\uparrow)}$   &$\displaystyle{\frac{G_{k_1}\times G_{k_4-k_1}}{G_{k_4}}\times G_{k_2-k_1}\times G_{k_3-k_2}\times G_{k_4-k_3}}$   &$\displaystyle{\frac{G_{k_1}\times G_{k_2-k_1}\times G_{k_3-k_2}\times G_{k_4-k_3}}{G_{k_4}}\times G_{k_4-k_1}}$\\
				\hline
				$\displaystyle{(\downarrow,\uparrow)}$ &$\displaystyle{\frac{G_{k_1}\times G_{k_4-k_1}}{G_{k_4}}\times\frac{G_{k_3}\times G_{k_2-k_3}}{G_{k_2}}\times G_{k_2-k_1}\times G_{k_4-k_3}}$  &$\displaystyle{\frac{G_{k_1}\times G_{k_2-k_1}}{G_{k_2}}\times\frac{G_{k_3}\times G_{k_4-k_3}}{G_{k_4}}\times G_{k_4-k_1}\times G_{k_2-k_3}}$ \\
				\hline
				$\displaystyle{(\uparrow,\downarrow)}$ &$\displaystyle{\frac{G_{k_1}\times G_{k_4-k_1}\times G_{k_3-k_4}}{G_{k_3}}\times G_{k_2-k_1}\times G_{k_3-k_2}}$ &$\displaystyle{\frac{G_{k_1}\times G_{k_2-k_1}\times G_{k_3-k_2}}{G_{k_3}}\times G_{k_4-k_1}\times G_{k_3-k_4}}$\\
				\hline
				$\displaystyle{(\downarrow,\downarrow)}$ &$\displaystyle{\frac{G_{k_1}\times G_{k_4-k_1}\times G_{k_3-k_4}\times G_{k_2-k_3}}{G_{k_2}}\times G_{k_2-k_1}}$ &$\displaystyle{\frac{G_{k_1}\times G_{k_2-k_1}}{G_{k_2}}\times G_{k_4-k_1}\times G_{k_3-k_4}\times G_{k_2-k_3}}$\\
				\hline
			\end{tabular}
		\end{center}
		\caption{The conformal field theories at the IR fixed points for $n=4$.}
		\label{table:ee'}
	\end{footnotesize}
\end{table}

\newpage
\noindent
\underline{$\bullet$ CFT at $(\downarrow,\uparrow)$}

\noindent
In this case we have four gauge fields and IR fixed point  is described by the action
\begin{equation}
\begin{split}\label{bb}
S=&S_{k_1}(a^{-1}_-g_1b_+)+S_{k_2}(b^{-1}_-g_2c_+)+S_{k_3}(c^{-1}_-g_3d_+)+S_{k_4}(d^{-1}_-g_4a_+)\\
&-S_{k_4}(a^{-1}_-a_+)-S_{k_2}(b^{-1}_-b_+)-S_{k_2}(c^{-1}_-c_+)-S_{k_4}(d^{-1}_-d_+)\\
&+S_{k_4-k_1}(a^{-1}_-)+S_{k_2-k_1}(b_+)+S_{k_2-k_3}(c^{-1}_-)+S_{k_4-k_3}(d_+)\, .
\end{split}
\end{equation}
This expression is found from \eqn{d} evaluated at the IR fixed point using the corresponding values in \eqn{upd4},
the relabeling $A_\pm^{(1)},\dots, A_\pm^{(4)}$ by $A_\pm,\dots , D_\pm$  and a parametrization of the latter like \eqn{i}. 
Integrating out the fields $(b_{\pm},d_{\pm})$, we find the following action
\begin{equation}
\begin{split}\label{cc}
S=&S_{k_1}(a^{-1}_-fc_+)+S_{k_2-k_1}(g_2c_+)+S_{k_3}(c^{-1}_-ha_+)+S_{k_4-k_3}(g_4a_+)\\
&+S_{k_4-k_1}(a^{-1}_-)+S_{k_2-k_3}(c^{-1}_-)-S_{k_4}(a^{-1}_-a_+)-S_{k_2}(c^{-1}_-c_+)\,,
\end{split}
\end{equation}
where $f=g_1g_2$ and $h=g_3g_4$. As before, the statement is that, \eqref{cc}  is the gauged fixed version of the action 
\be
\begin{split}
\label{dd}
S_{G/H}=&S_{k_1}(a^{-1}_-g_1c_+)+S_{k_3}(c^{-1}_-g_2a_+)+S_{k_2-k_1}(g_3c_+)+S_{k_2-k_3}(c^{-1}_-g_4)\\
&+S_{k_4-k_3}(g_5a_+)+S_{k_4-k_1}(a^{-1}_-g_6)-S_{k_4}(a^{-1}_-a_+)-S_{k_2}(c_-^{-1}c_+)\, ,
\end{split}
\end{equation}
with the gauge choice $g_4=g_6=\mathbb{1}$.
The gauge symmetries of \eqref{dd} are clearly
\begin{equation}\label{gs}
\begin{split}
& (g_1,g_2,g_3,g_4,g_5,g_6) \to (h_1^{-1}g_1h_2,h^{-1}_2g_2h_1,g_3h_2,h^{-1}_2g_4,g_5h_1,h^{-1}_1g_6)\ ,
\\
&(a_{\pm},c_{\pm})\to(h_1^{-1}a_{\pm},h_2^{-1}c_{\pm})\,,
\end{split}
\end{equation}
where $h_1(\s_+,\s_-), h_2(\s_+,\s_-)\in G$, while the chiral ones are given by
\begin{equation}\label{cs}
(g_3,g_4,g_5,g_6)\to(\Omega^{-1}_L(\s_+)g_3,g_4\Omega_R(\s_-),\tilde{\Omega}^{-1}_L(\s_+)g_5,g_6\tilde{\Omega}_R(\s_-))\,.
\end{equation}
The symmetries in \eqref{gs}, \eqref{cs} exactly correspond to the symmetries presented in the second row of table \ref{table:ee'}.\par
\noindent
\underline{$\bullet$ CFT at $(\uparrow,\downarrow)$}\par
\noindent
The corresponding CFT is described by the action
\begin{equation}\label{ee}
\begin{split}
S&=S_{k_1}(a^{-1}_-g_1b_+)+S_{k_2}(b^{-1}_-g_2c_+)+S_{k_3}(c^{-1}_-g_3d_+)+S_{k_4}(d^{-1}_-g_4a_+)\\
&-S_{k_4}(a^{-1}_-a_+)-S_{k_2}(b^{-1}_-b_+)-S_{k_3}(c^{-1}_-c_+)-S_{k_3}(d^{-1}_-d_+)\\
&+S_{k_4-k_1}(a^{-1}_-)+S_{k_2-k_1}(b_+)+S_{k_3-k_2}(c_+)+S_{k_3-k_4}(d^{-1}_-)\, ,
\end{split}
\end{equation}
which is found from \eqn{d} evaluated at the IR fixed point using \eqn{upd4}.
Integrating over the fields $(a_{\pm},b_{\pm},c_{\pm})$ we find that the action becomes
\begin{equation}\label{ff}
\begin{split}
S=&S_{k_1}(d^{-1}_-fd_+)+S_{k_2-k_1}(hd_+)+S_{k_3-k_2}(g_3d_+)+\\
&+S_{k_4-k_1}(d^{-1}_-g_4)+S_{k_3-k_4}(d^{-1}_-)-S_{k_3}(d^{-1}_-d_+)\,,
\end{split}
\end{equation}
where $f=g_4g_1g_2g_3$ and $h=g_2g_3$.
The above action can be viewed as the gauged fixed version, $g_5=\mathbb{1}$, of the following gauged action
\begin{equation}\label{gg}
\begin{split}
S_{G/H}=&S_{k_1}(d^{-1}_-fd_+)+S_{k_2-k_1}(hd_+)+S_{k_3-k_2}(g_3d_+)+\\
&+S_{k_4-k_1}(d^{-1}_-g_4)+S_{k_3-k_4}(d^{-1}_-g_5)-S_{k_3}(d^{-1}_-d_+)\,.
\end{split}
\end{equation}
Identifying the gauge and chiral symmetries of \eqref{gg}, we see that they correspond exactly to the ones presented in the third row of table \ref{table:ee'}.
Indeed,  the gauge symmetry of \eqn{gg} is
\begin{equation}
H:(g_1,g_2,g_3,g_4,g_5)\to (h^{-1}g_1h,g_2h,g_3h,h^{-1}g_4,h^{-1}g_5),\quad d_{\pm}\to h^{-1}d_{\pm} \,,
\end{equation}
where $h(\s_+,\s_-)\in G$ , whereas the chiral algebra symmetry is
\begin{equation}
(g_2,g_3,g_4,g_5)\to (\Omega_L^{-1}(\s_+)g_2,\tilde{\Omega}_L^{-1}g_3,g_3\Omega_R^{-1}(\s_-),g_4\tilde{\Omega}_R^{-1}(\s_-))\ .
\end{equation}

\no
Notice  that, the CFTs at the IR fixed points $(\downarrow,\uparrow)$, $(\uparrow,\downarrow)$ are selfdual under the
generalized parity transformation and the renaming $k_2\leftrightarrow k_4$, in the sense that the symmetry of the left sector is mapped to the symmetry of the right sector and vice versa (see table \ref{table:ee'}).  
In contrast, the CFTs at the IR fixed points $(\uparrow,\uparrow)$, $(\downarrow,\downarrow)$ are mapped to each other 
under the generalized parity transformation and the same renaming as before $k_2\leftrightarrow k_4$.

\subsubsection{Remarks on the conformal symmetry for arbitrary $n$}

We conclude this subsection with some remarks for the general case of $n$ coupled models.
We will not present the intermediate details which are a more complicated version of the low $n$ cases we have already analyzed.
 
\no
Consider the basic ordering of levels  $k_1<k_2<\dots<k_n$. The current algebra part of the CFT for both sectors is \eqn{o},
whereas the full conformal symmetry is
\begin{equation}
\begin{split}
&\frac{G_{k_1}\times G_{k_n-k_1}}{G_{k_n}}\times G_{k_2-k_1}\times\dots\times G_{k_{n-1}-k_{n-2}} \times G_{k_n-k_{n-1}}\biggr|_{L}\ ,
\\
&\frac{G_{k_1}\times G_{k_2-k_1}\times\dots \times G_{k_{n-1}-k_{n-2}} \times G_{k_n-k_{n-1}}}{G_{k_n}}\times G_{k_n-k_{1}}\biggr|_{R}\ .
\end{split}
\end{equation}
A slightly changed ordering of the levels, for example that with $k_i<k_{i-1}$ for $i=2,3,\dots , n-1$ led to  the current algebra part of the CFT found in \eqn{o'}.
The full conformal symmetry in that case is
\ba
\label{gggg}
&&\frac{G_{k_1}\times G_{k_n-k_1}}{G_{k_n}}\times\frac{G_{k_i}\times G_{k_{i-1}-k_i}}{G_{k_{i-1}}}\times G_{k_2-k_1}\times\dots\times  \hat{G}_{k_{i}-k_{i-1}}\times\dots
\times  G_{k_n-k_{n-1}}\biggr|_{L}\ ,
\\
&&\frac{G_{k_1}\times G_{k_2-k_1}\times\dots \times G_{k_{i-1}-k_{i-2}}}{G_{k_{i-1}}}\times\frac{G_{k_i}\times G_{k_{i+1}-k_i}\times\dots\times G_{k_{n}-k_{n-1}}}{G_{k_{n}}}\times G_{k_{i-1}-k_{i}}\times G_{k_{n}-k_1}\biggr|_{R}\ ,
\nonumber
\ea
where we note that the hat in $G$ in the line corresponding to the left sector implies that  this term is missing and instead it appears in second coset with reverse sign for the level which makes it positive.  
Note also that the case with $i=1$ cannot be realized since $k_1$ is chosen to be the smallest level.
The case with  $i=n$ is also special and there is a modification in \eqn{gggg} above given by
\begin{equation}
\label{ggg}
\begin{split}
&\frac{G_{k_1}\times G_{k_n-k_1}\times G_{k_{n-1}-k_n}}{G_{k_{n-1}}}\times G_{k_2-k_1}\times\dots \times G_{k_{n-2}-k_{n-3}}\times G_{k_{n-1}-k_{n-2}}\biggr|_{L}\ ,
\\
&\frac{G_{k_1}\times G_{k_2-k_1}\times\dots\times G_{k_{n-2}-k_{n-3}}\times G_{k_{n-1}-k_{n-2}}}{G_{k_{n-1}}}\times G_{k_{n-1}-k_n} \times G_{k_n-k_1} \biggr|_{R}\ .
\end{split}
\end{equation}

We next present the  general rule for identifying the IR fixed point CFTs. It turns out that it is more convenient to discuss this in terms
of the  $\l_0^{(i)}$'s instead of the ordering of the levels as we did in the previous two examples. 
Let us begin with a fixed but otherwise arbitrary IR fixed point which in our notation is an array of arrows pointing upwards or 
donwards. In that array, for each $\l_0^{(i)}$ term, corresponding to a pointing upwards
arrow $\uparrow_i$, we write down a chiral algebra symmetry in the left sector and a coset type symmetry in the right sector. 
In contrast, for each $(\l_0^{(i)})^{-1}$ term, corresponding to a pointing downwards
arrow $\downarrow_i$, a coset type in the left sector and a chiral algebra symmetry in the right sector is written.
Thus, in detail we have that
\be
\begin{split}
\label{rrrr}
&\uparrow_i\quad \Longrightarrow\quad  G_{k_{i}-k_{i-1}}\biggr|_{L}\!\times \frac{G_{k_{i-1}}\times G_{k_{i}-k_{i-1}}}{G_{k_{i}}}\biggr|_{R}\! ,
\\
&
\downarrow_i   \quad\Longrightarrow\quad \frac{G_{k_{i}}\times G_{k_{i-1}-k_{i}}}{G_{k_{i-1}}}\biggr|_{L}\times G_{k_{i-1}-k_{i}}\biggr|_{R}\!\,.
\end{split}
\ee
The above rule applies solely when no two adjacent arrows in the array point in the same direction.
A refinement of this rule is needed if this is not the case. We present it first for the case when two adjacent arrows are pointing in 
the same direction either upwards or downwards
\be
\begin{split}
\label{rrrrr}
                &\uparrow_i\, ,  \uparrow_{i+1}
                 \quad \Longrightarrow\quad  G_{k_{i}-k_{i-1}}\times G_{k_{i+1}-k_{i}}\biggr|_{L}\times \frac{G_{k_{i-1}}\times G_{k_{i}-k_{i-1}}\times G_{k_{i+1}-k_{i}}}{G_{k_{i+1}}}\biggr|_{R}\! ,
                \\
                &\downarrow_i\, ,  \downarrow_{i+1}
                   \quad\Longrightarrow\quad \frac{G_{k_{i+1}}\times G_{k_{i}-k_{i+1}}\times G_{k_{i-1}-k_{i}}}{G_{k_{i-1}}}\biggr|_{L}\times G_{k_{i-1}-k_{i}}\times G_{k_{i}-k_{i+1}}\biggr|_{R} \! .
\end{split}
\ee     
Obviously, this rule is extentable for more adjacent arrows pointing in the same direction.
Also, we should keep in mind that the array of arrows forms a closed chain, implying that the 1st and the $n$th should be consider as adjacent ones.

\no
For the reader's convenience we present an explicit example corresponding to the $n=6$ case with array
\begin{equation}\label{example}
(\downarrow_1,\uparrow_2,\downarrow_3,\uparrow_4,\downarrow_5,\downarrow_6) \ .
\end{equation}
Using the above rules we have that
\begin{equation}
\begin{split}\label{exam}
&\frac{G_{k_1}\times G_{k_6-k_1}\times G_{k_5-k_6}\times G_{k_4-k_5}}{G_{k_4}}\times\frac{G_{k_3}\times G_{k_2-k_3}}{G_{k_2}}\times G_{k_2-k_1}\times G_{k_4-k_3}\biggr|_{L}\ ,
\\
&\frac{G_{k_1}\times G_{k_2-k_1}}{G_{k_2}}\times\frac{G_{k_3}\times G_{k_4-k_3}}{G_{k_4}}\times G_{k_2-k_3}\times G_{k_6-k_1}\times G_{k_5-k_6}\times G_{k_4-k_5}\biggr|_{R}.
\end{split}
\end{equation}
Notice that in \eqref{example}, due to the cyclicity of our models, there are three consecutive arrows pointing downwards, those at positions 5, 6 and 1. This is the origin of the first coset in the first line of 
\eqref{exam}.
 Finally, the fact that the levels in all coset-type symmetries sum up to zero is very helpful when applying the aforementioned rules.

\subsection{Independent IR CFTs}

As already mentioned, for every $n$ the transformation \eqn{sysmmmr} involving the renaming of the index $i$ and the parity transformation leaves the action \eqref{d} invariant.
In the weak coupling limit, it is a transformation that does not change the interaction of the nearest neighbors in \eqref{d1}.
We rewrite here the part of this symmetry transformation needed for our purposes below
\begin{equation}
\label{3.55}
\mathcal{R}_n:\quad A_\pm^{(i)}\to A_\mp^{(n+3-i)}\, ,
\ k_i\to k_{n+2-i}\, , \  k^{(i)}\to k^{(n+3-i)}\, , \ \l_0^{(i)}\to \big(\l_0^{(n+3-i)}\big)^{-1}\, .
\end{equation}
In \eqref{3.55} we have also included  \eqn{sysmmmr1} which will be useful when studying the CFTs at the IR fixed points.
Our aim here is to discover which of the previously identified $2^{n-2}$ 
IR fixed point CFTs are in fact the independent ones. 
Towards that, an elegant way to realize this symmetry pictorially with the aid of a polygon will be most useful. 
We assign to the $i$'th vertex of an $n$-polygon the $i$'th interaction term of the fields $A^{(i)}_{\pm}$, and to every side connecting the edges $i$ and $i+1$ the level $k_i$, which is the prefactor of the $i$'th propagating term in \eqref{d} connecting the fields $A^{(i)}_-$ and $A^{(i+1)}_+$. Doing so, the transformation $\mathcal{R}_n$ is realized as the reflection about the perpendicular bisector through the side $k_1$, denoted with the blue line in the following figures. In figure \eqref{fig:generaltriangle} we give an example for the case of $n=3$ where the action \eqref{d} is represented as a triangle.
\noindent 
For the reader's convenience we give also the Lagrangian, not including for the sake of simplicity the WZW terms
\begin{align}
\label{action3}
\nonumber
L&=k_1\Tr(A_-J^{(1)}_+-B_+J^{(1)}_-+A_-g_1B_+g_1^{-1})+k_2\Tr(B_-J^{(2)}_+-C_+J^{(2)}_-+B_-g_2C_+g_2^{-1})\\\nonumber
&\quad +k_3\Tr(C_-J^{(3)}_+-A_+J^{(3)}_-+C_-g_3A_+g_3^{-1})
-k^{(1)}\l^{-1}_1\Tr(A_+A_-)-k^{(2)}\l^{-1}_2\Tr(B_+B_-)\\
&\quad -k^{(3)}\l^{-1}_3\Tr(C_+C_-)\,.
\end{align}
\begin{figure}[h]
\centering
\includegraphics[scale=1.5]{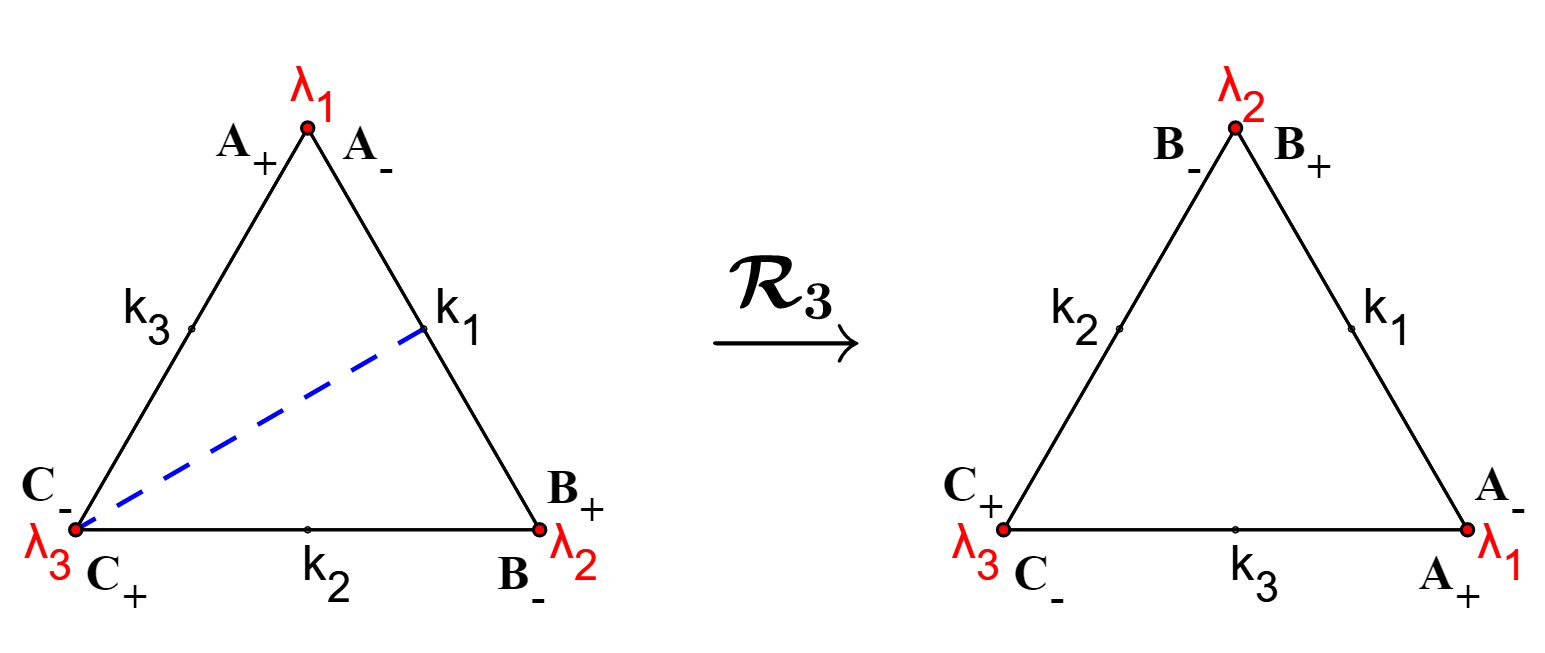}
\caption{\footnotesize{The vertices of the right triangle represent the three last terms of \eqref{action3}, where for 
simplicity we did not include the $k^{(i)}$ terms and we assigned the $\l_i$ value instead of its inverse. Every side of the triangle, which connects the appropriate fields, is denoted with the appropriate level $k_i$ which appears as the prefactor of the corresponding propagating term from the first and second line of \eqref{action3}. One can check that $\mathcal{R}_3$ transforms the fields of \eqref{action3} as $J^{(1)}_+\leftrightarrow -J^{(1)}_-\,,J^{(2)}_{\pm}\leftrightarrow -J^{(3)}_{\mp}\,,A_{\pm}\leftrightarrow B_{\mp}\,,C_{\pm}\to C_{\mp}$ the deformation parameters as $\l_1\leftrightarrow\l_2$, $\l_3\to\l_3$ and the levels as $k_1\to k_1$ and $k_2\leftrightarrow k_3$. Thus in the framework of polygons $\mathcal{R}_3$ transforms the left triangle to the right one.}}
\label{fig:generaltriangle}
\end{figure}
\no
Next we would like to understand the action of the symmetry \eqn{3.55} at the IR fixed point CFTs.
In order to do this we replace each $\l_i$ at the edges of the polygon with the corresponding arrow, $\uparrow_i$ or $\downarrow_i$. 
Under $\mathcal{R}_n$, each arrow will be reversed, if the reflected arrows with respect to the bisector are pointing to the same direction
(both are up or down) and it will remain the same if they are opposite. This rule originates from the transformation of the $\l_0^{(i)}$ in \eqref{3.55}.
Seemingly different fixed points related with the above transformation describe equivalent CFTs.

\noindent 
When $n$ is odd, the independent CFTs are $\frac{2^{n-2}}{2}$, that is 
\be
{\rm For}\ n\ {\rm odd}:\qq 2^{n-3}\ ,
\ee 
since every fixed point has a dual corresponding to a different array of arrows as the perpendicular bisector always passes from a vertex of the polygon. Thus according to \eqref{3.55} the arrow of this vertex will be  reversed under $\mathcal{R}_n$. 

\no
For the case of n even things are a bit more involved as in this case there exist IR fixed points which are self dual. Self dual IR points, are the points, in which the arrows of the edges that are mirror images of each other with respect to the perpendicular bisector have opposite directions. We find that the independent CFTs are 
$\frac{2^{n-2}-2^{\frac{n-2}{2}}}{2}+2^{\frac{n-2}{2}}$, that is 
\be
{\rm For}\ n\ {\rm even}:\qq 2^{n-3}+2^{\frac{n}{2}-2}\ .
\ee 
Notice that,  for $n$ odd there are no self dual fixed points.

\no
We  now turn our attention to the cases presented in the previous sections, namely those with $n=3,4$ and then we also consider  the $n=5,6$ cases where the pictorial method we developed is particularly very useful.  

\begin{figure}[t]
\centering
\includegraphics[scale=1.1]{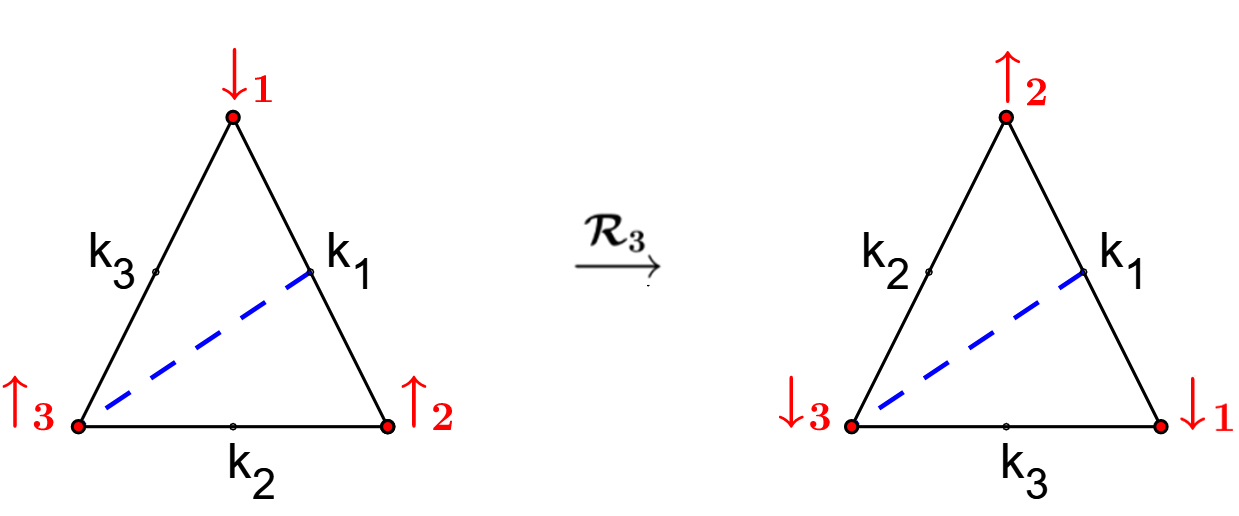}
\caption{\footnotesize{The left triangle represents the CFT defined at the IR point $(\uparrow)$ and the right triangle the CFT at the IR point $(\downarrow)$}.}
\label{fig:triangle}
\end{figure}

\noindent 
Consider first the case with $n=3$. 
The transformation $\mathcal{R}_3$, in the framework of polygons, is
\begin{equation}
\label{123}
\mathcal{R}_3:\qquad 1\leftrightarrow 2\, , \quad    k_2\leftrightarrow k_3\, ,\quad 
\l_0^{(1)} \leftrightarrow (\l_0^{(2)})^{-1}\ ,
\end{equation}
where for simplicity we replaced the pair of gauge fields $A^{(i)}_{\pm}$ at every vertex with the label $i$ of the vertex, a notation we will use subsequently. In addition, we omit writing indices and quantities that remain invariant under the 
symmetry transformation. 
The right triangle in fig. \eqref{fig:triangle} corresponds to the CFT defined at the IR point denoted as $(\uparrow)$. 
Since points $1$ and $2$ are mirror images of each other with respect to the blue line and the corresponding arrows have opposite directions, they do not change under $\mathcal{R}_3$, while the arrow at point $3$ is reversed. The resulting set up is the right triangle in fig. \ref{fig:triangle} which corresponds to the CFT at the IR point denoted as $(\downarrow)$. 
We conclude that the symmetry \eqn{123} relates the two CFTs in table \eqref{table:eeee'} and that there is only one independent CFT. 

\begin{figure}[h]
\centering
\includegraphics[scale=1.1]{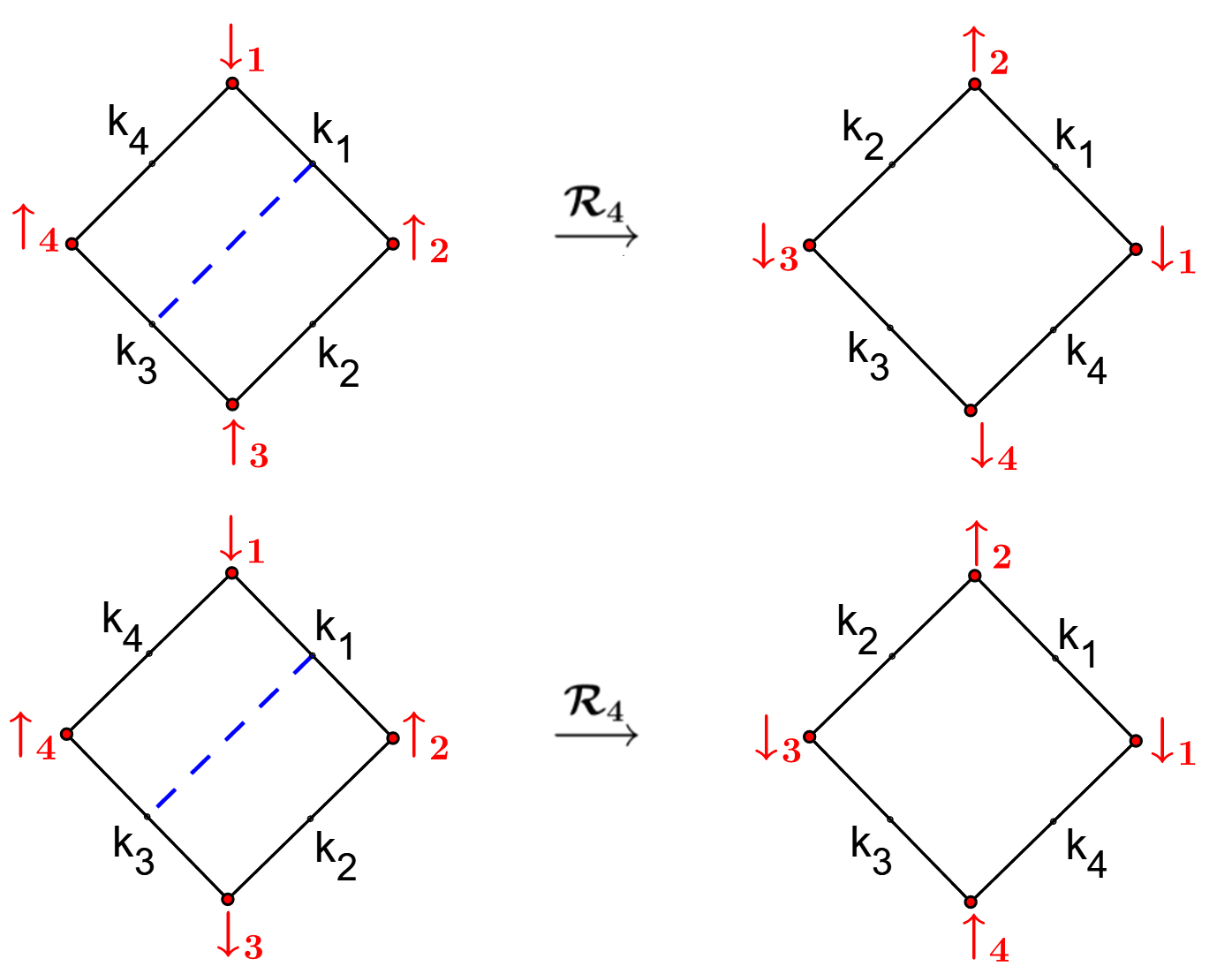}
\caption{\footnotesize{The first line corresponds to the dual CFTs defined at the IR points $(\uparrow,\uparrow)$, $(\downarrow,\downarrow)$ and the second line to the selfdual CFT  at $(\downarrow,\uparrow)$}.}
\label{fig:squarex}
\end{figure}

\noindent 
For the case of $n=4$, the symmetry operation $\mathcal{R}_4$ acts as
\begin{equation}
\label{357}
\mathcal{R}_4:\qq 1\leftrightarrow 2\, , \quad 3\leftrightarrow 4\,  ,\quad k_2\leftrightarrow k_4\, ,\quad 
\l_0^{(1)} \leftrightarrow (\l_0^{(2)})^{-1}\, , \quad \l_0^{(3)} \leftrightarrow (\l_0^{(4)})^{-1} \, .
\end{equation}
In figure \eqref{fig:squarex} we present the relation between the CFTs at the IR points $(\uparrow,\uparrow)$, $(\downarrow,\downarrow)$ and the self duality of the fixed point  $(\downarrow,\uparrow)$  under \eqref{357}. We did not include the self duality of the IR point $(\uparrow,\downarrow)$ as it is straightforward in this framework. Hence, 
among  the CFTs presented in table \eqref{table:ee'} there are three independent ones.   

\begin{figure}[h!]
	\centering
	\includegraphics[scale=1.1]{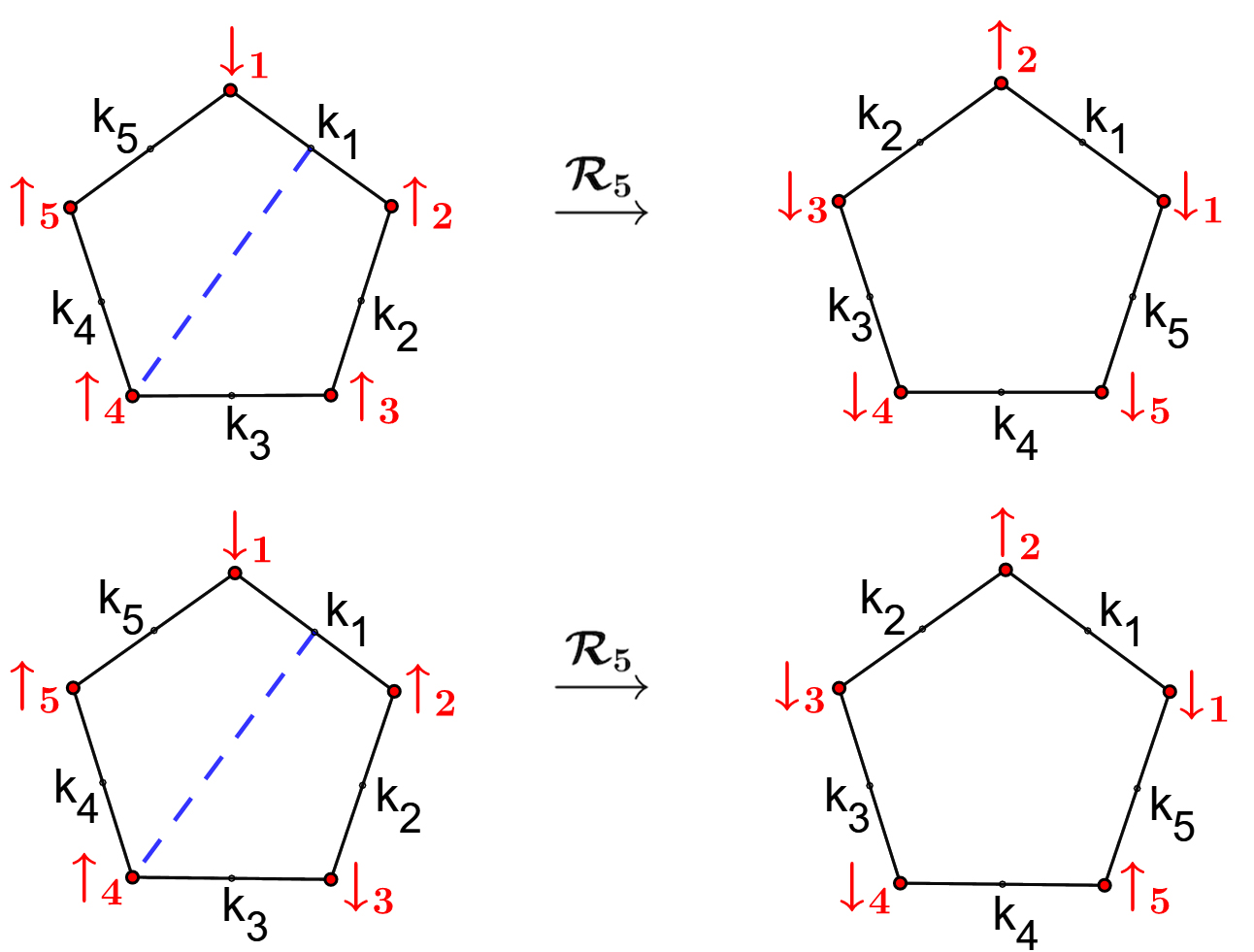}
\caption{\footnotesize{The first and second line correspond to the dual CFTs at the IR points $(\uparrow,\uparrow,\uparrow)$, $(\downarrow,\downarrow,\downarrow)$ and $(\downarrow,\uparrow,\uparrow)$, $(\downarrow,\downarrow,\uparrow)$, respectively.}}
	\label{fig:pentagon}
\end{figure}

\no
In the case with $n=5$ there are eight different IR fixed points CFTs which are presented in table \eqref{table:eeeee'}. 
The symmetry acts as 
\begin{equation}
\begin{split}
\label{ssym5}
\mathcal{R}_5:\qquad & 1\leftrightarrow 2\, , \quad 3\leftrightarrow 5\,  ,\quad k_2\leftrightarrow k_5\, ,\quad
 k_3\leftrightarrow k_4\, ,
 \\
 &
\l_0^{(1)} \leftrightarrow (\l_0^{(2)})^{-1}\, , \quad \l_0^{(3)} \leftrightarrow (\l_0^{(5)})^{-1} \, .
\end{split}
\end{equation}
\no
In figure \eqref{fig:pentagon} we present the related CFTs at the IR points $(\uparrow,\uparrow,\uparrow)$, $(\downarrow,\downarrow,\downarrow)$ and $(\uparrow,\downarrow,\uparrow)$, $(\downarrow,\uparrow,\downarrow)$ as polygons. 

\no
A more direct but equivalent way towards finding the equivalent IR points under $\mathcal{R}_5$ is the following. 
In a given array of arrows denoting the IR fixed point CFT, we compare the first one with the last one, the second with the next to the last and so on and so forth. If they are pointing in the same direction we reverse their direction, whereas if they point in opposite directions we keep these the same. For the middle one, for odd values of $n$ as in this example, we reverse its direction.

\no
Thus we have the equivalent IR fixed  point CFTs
\begin{equation}
\begin{split}
&(\uparrow,\uparrow,\uparrow)\sim(\downarrow,\downarrow,\downarrow)\,, \qq (\downarrow,\uparrow,\uparrow)\sim(\downarrow,\downarrow,\uparrow)\,,
\\
&(\uparrow,\downarrow,\uparrow)\sim(\downarrow,\uparrow,\downarrow)\,, \qq
(\uparrow,\uparrow,\downarrow)\sim(\uparrow,\downarrow,\downarrow)\, .
\\
\end{split}
\end{equation}
Thus for $n=5$ there are four independent CFTs among the entries of table \eqref{table:eeeee'}.

\no
Finally in the $n=6$ case there are sixteen different IR fixed CFT points. According to  
 our analysis, only ten of them define independent ones.  The equivalent ones related by $\mathcal{R}_6$ are
\begin{equation}
\begin{split}
&(\uparrow,\uparrow,\uparrow,\uparrow)\sim(\downarrow,\downarrow,\downarrow,\downarrow)\quad,\quad(\downarrow,\uparrow,\uparrow,\downarrow)\sim(\uparrow,\downarrow,\downarrow,\uparrow)\,,\\
&(\uparrow,\uparrow,\uparrow,\downarrow)\sim(\uparrow,\downarrow,\downarrow,\downarrow)\quad,\quad(\downarrow,\downarrow,\uparrow,\downarrow)\sim(\uparrow,\downarrow,\uparrow,\uparrow)\,,\\
&(\downarrow,\downarrow,\downarrow,\uparrow)\sim(\downarrow,\uparrow,\uparrow,\uparrow)\quad,\quad(\downarrow,\uparrow,\downarrow,\downarrow)\sim(\uparrow,\uparrow,\downarrow,\uparrow)\, ,
\end{split}
\end{equation}
whereas the selfdual ones are $(\downarrow,\downarrow,\uparrow,\uparrow),(\downarrow,\uparrow,\downarrow,\uparrow),(\uparrow,\downarrow,\uparrow,\downarrow),(\uparrow,\uparrow,\downarrow,\downarrow)$.

\begin{table}[p]
	\begin{footnotesize}
		\renewcommand{\arraystretch}{2.7}
		\begin{center}
			\begin{tabular}{ |p{1.1cm}||p{8.7cm}|}
				\hline
				IR     & Left sector\\
				\hline
				$\displaystyle{(\uparrow,\uparrow,\uparrow)}$   &$\displaystyle{\frac{G_{k_1}\times G_{k_5-k_1}}{G_{k_5}}\times G_{k_2-k_1}\times G_{k_3-k_2}\times G_{k_4-k_3}\times G_{k_5-k_4}}$\\
				\hline
				$\displaystyle{(\downarrow,\uparrow,\uparrow)}$ &$\displaystyle{\frac{G_{k_1}\times G_{k_5-k_1}}{G_{k_5}}\times\frac{G_{k_3}\times G_{k_2-k_3}}{G_{k_2}}\times G_{k_2-k_1}\times G_{k_4-k_3}\times G_{k_5-k_4}}$ \\
				\hline
				$\displaystyle{(\uparrow,\downarrow,\uparrow)}$ &$\displaystyle{\frac{G_{k_1}\times G_{k_5-k_1}}{G_{k_5}}\times\frac{G_{k_4}\times G_{k_3-k_4}}{G_{k_3}}\times G_{k_2-k_1}\times G_{k_3-k_2}\times G_{k_5-k_4}}$\\
				\hline
				$\displaystyle{(\uparrow,\uparrow,\downarrow)}$ &$\displaystyle{\frac{G_{k_1}\times G_{k_5-k_1}\times G_{k_4-k_5}}{G_{k_4}}\times G_{k_2-k_1}\times G_{k_3-k_2}\times G_{k_4-k_3}}$\\
				\hline
				$\displaystyle{(\downarrow,\downarrow,\uparrow)}$ &$\displaystyle{\frac{G_{k_1}\times G_{k_5-k_1}}{G_{k_5}}\times\frac{G_{k_4}\times G_{k_3-k_4}\times G_{k_2-k_3}}{G_{k_2}}\times G_{k_2-k_1}\times G_{k_5-k_4}}$\\
				\hline
				$\displaystyle{(\downarrow,\uparrow,\downarrow)}$ &$\displaystyle{\frac{G_{k_3}\times G_{k_2-k_3}}{G_{k_2}}\times\frac{G_{k_1}\times G_{k_5-k_1}\times G_{k_4-k_5}}{G_{k_4}}\times G_{k_2-k_1}\times G_{k_4-k_3}}$\\
				\hline
				$\displaystyle{(\uparrow,\downarrow,\downarrow)}$ &$\displaystyle{\frac{G_{k_1}\times G_{k_5-k_1}\times G_{k_4-k_5}\times G_{k_3-k_4}}{G_{k_3}}\times G_{k_2-k_1}\times G_{k_3-k_2}}$\\
				\hline
				$\displaystyle{(\downarrow,\downarrow,\downarrow)}$   &$\displaystyle{\frac{G_{k_1}\times G_{k_5-k_1}\times G_{k_4-k_5}\times G_{k_3-k_4}\times G_{k_2-k_3}}{G_{k_2}}\times G_{k_2-k_1}}$\\
				\hline
			\end{tabular}
		\vspace{9mm}
		\begin{tabular}{ |p{1.1cm}||p{8.7cm}|}
			\hline
			IR     & Right sector\\
			\hline
			$\displaystyle{(\uparrow,\uparrow,\uparrow)}$    &$\displaystyle{\frac{G_{k_1}\times G_{k_2-k_1}\times G_{k_3-k_2}\times G_{k_4-k_3}\times G_{k_5-k_4}}{G_{k_5}}\times G_{k_5-k_1}}$\\
			\hline
			$\displaystyle{(\downarrow,\uparrow,\uparrow)}$ &$\displaystyle{\frac{G_{k_1}\times G_{k_2-k_1}}{G_{k_2}}\times\frac{G_{k_3}\times G_{k_4-k_3}\times G_{k_5-k_4}}{G_{k_5}}\times G_{k_5-k_1}\times G_{k_2-k_3}}$ \\
			\hline
			$\displaystyle{(\uparrow,\downarrow,\uparrow)}$ &$\displaystyle{\frac{G_{k_4}\times G_{k_5-k_4}}{G_{k_5}}\times\frac{G_{k_1}\times G_{k_2-k_1}\times G_{k_3-k_2}}{G_{k_3}}\times G_{k_5-k_1}\times G_{k_3-k_4}}$\\
			\hline
			$\displaystyle{(\uparrow,\uparrow,\downarrow)}$ &$\displaystyle{\frac{G_{k_1}\times G_{k_2-k_1}\times G_{k_3-k_2}\times G_{k_4-k_3}}{G_{k_4}}\times G_{k_5-k_1}\times G_{k_4-k_5}}$\\
			\hline
			$\displaystyle{(\downarrow,\downarrow,\uparrow)}$ &$\displaystyle{\frac{G_{k_1}\times G_{k_2-k_1}}{G_{k_2}}\times\frac{G_{k_4}\times G_{k_5-k_4}}{G_{k_5}}\times G_{k_5-k_1}\times G_{k_3-k_4}\times G_{k_2-k_3}}$\\
			\hline
			$\displaystyle{(\downarrow,\uparrow,\downarrow)}$ &$\displaystyle{\frac{G_{k_1}\times G_{k_2-k_1}}{G_{k_2}}\times\frac{G_{k_3}\times G_{k_4-k_3}}{G_{k_4}}\times G_{k_5-k_1}\times G_{k_4-k_5}\times G_{k_2-k_3}}$\\
			\hline
			$\displaystyle{(\uparrow,\downarrow,\downarrow)}$ &$\displaystyle{\frac{G_{k_1}\times G_{k_2-k_1}\times G_{k_3-k_2}}{G_{k_3}}\times G_{k_5-k_1}\times G_{k_4-k_5}\times G_{k_3-k_4}}$\\
			\hline
			$\displaystyle{(\downarrow,\downarrow,\downarrow)}$   &$\displaystyle{\frac{G_{k_1}\times G_{k_2-k_1}}{G_{k_2}}\times G_{k_5-k_1}\times G_{k_4-k_5}\times G_{k_3-k_4}\times G_{k_2-k_3}}$\\
			\hline
		\end{tabular}
		\end{center}
		\caption{Left and right sector of the CFTs for $n=5$.}
		\label{table:eeeee'}
	\end{footnotesize}
\end{table}

\section{Concluding remarks}
We have constructed a large class of integrable multiparametric deformations of $n$ WZW $\s$-models each at a different level $k_i$. The theories considered  can be rewritten as a special case of the most genaral $\l$-deformed models of \cite{Georgiou:2018gpe} but they were not explicitly considered in that work. Our models flow from a CFT which at the UV point has the  symmetry group $G_{k_1}\times G_{k_2}\times\dots\times G_{k_n}$  to different IR fixed points depending on the number $n$ and on the different orderings of the levels $k_i$. This happens because in our construction the WZW models are cyclicly coupled, similarly to \cite{Georgiou:2017oly,Georgiou:2017jfi,Georgiou:2016urf}. The goal of this paper was to classify the resulting IR CFTs.

In order to determine the conformal symmetry of each IR CFT, we constructed gauge invariant actions which, after fixing the gauge, describe the corresponding IR CFTs. The subgroup gauged at each case, is a different, anomaly free, subgroup of the global $G_L\times G_R$ symmetry of the $n$ WZW models. 
This procedure lead to the conclusion that the left and right sector of each of the IR CFTs is based on different products of coset and affine type  conformal symmetries. Despite this asymmetry between sectors, the left and right central charge are the same and in agreement with the central charge read from the exact in the deformation parameters C-function which we also have derived. Furthermore, we have shown that there are CFTs defined at different fixed points which are related by a transformation,  the parity transformation \eqref{3.55}. That allowed for the final classification of the
 inequivalent  IR fixed point CFTs.      
    
An interesting future direction is to apply the formalism developed in this work in order to characterize and determine the symmetries of the IR fixed points to which the models constructed in \cite{Georgiou:2018hpd} and \cite{Georgiou:2018gpe} flow. These models have an explicit Lagrangian realization and their RG flow equations have a much richer structure that the theories considered here. In that respect it would be important to see what modifications, if any, will be needed in achieving the characterization of their IR fixed points. Furthermore, it would be also interesting to study the RG flow equations and their fixed points of the integrable models of \cite{Bassi:2019aaf} (see also \cite{Delduc:2019bcl}). It would also be important to examine if the integrable models presented in this work, as well as the ones in \cite{Georgiou:2017oly,Georgiou:2018hpd,Georgiou:2018gpe}, belong to a class of the E -models of \cite{Klimcik:2015gba}. This may enable one to identify the corresponding integrable $\eta$-deformed models introduced in \cite{Klimcik:2002zj,Klimcik:2008eq,Klimcik:2014} for the case of group spaces. 
Finally, it would be interesting to determine the integrability preserving D-branes all the way along the RG flow \eqref{e} of our models. Our theories posses non-trivial IR fixed points,  in contradistinction to the work of \cite{Driezen:2018glg} which considered this problem for the prototype deformed 
models of \cite{Sfetsos:2013wia} which possess no such points.   
This renders the embedding of D-branes in the target space a more involved problem.

\subsection*{Acknowledgments}
The work of G.G. on this project has received funding from the Hellenic Foundation for Research and Innovation
(H.F.R.I) and the General Secretariat for Research and Technology (GSRT), under grant
agreement No 15425. \\
The research work of K. Sfetsos was supported by the Hellenic Foundation for 
Research and Innovation (H.F.R.I.) under the ``First Call for H.F.R.I.
Research Projects to support Faculty members and Researchers and
the procurement of high-cost research equipment grant'' (MIS 1857, Project Number: 16519). \\
The research of G. Pappas is co-financed by Greece and the European Union (European Social Fund- ESF) through the Operational Programme Human Resources Development, Education and Lifelong Learning in the context of the project ``Strengthening Human Resources Research Potential via Doctorate Research'' (MIS-5000432), implemented by the State Scholarships Foundation (I.K.Y).

\appendix
\section{Integrability}

In this appendix we will show classical integrability of the $\s$-model action $\eqref{d}$, for the isotropic case, by determining the Lax pairs explicitly.
Varying the action with respect to the gauge fields we get the following constraints
\begin{equation}\label{3}
\begin{split}
& \l^{-1}_i\l_0^{(i)}A^{i}_+=\partial_+g_ig^{-1}_i+g_iA^{(i+1)}_+g^{-1}_i\,,\\
&\l^{-1}_{i}(\l_0^{({i})})^{-1}A^{i}_-=-g^{-1}_{i-1}\partial_-g_{i-1}+g^{-1}_{i-1}A^{(i-1)}_-g_{i-1}\,.
\end{split}
\end{equation}
Defining the covariant derivative
\begin{equation}\label{4}
D_{\pm}g_i=\partial_{\pm}g_i-A^{i}_{\pm}g_i+g_iA^{(i+1)}_{\pm}\,,
\end{equation}
the equation \eqref{3} can be rewritten as
\begin{equation}\label{5}
\begin{split}
&D_+g_ig^{-1}_i=\Big(\l^{-1}_i\l_0^{(i)}-1\Big)A^{(i)}_+,\\\,
&g^{-1}_{i-1}D_+g_{i-1}=-\Big(\l^{-1}_{i}(\l_0^{(i)})^{-1}-1\Big)A^{(i)}_-\,.
\end{split}
\end{equation}
Varying again the action $\eqref{d}$ with respect to $g_i$ and using the definition \eqref{4} we obtain that
\begin{equation}\label{6}
\begin{split}
&D_-(D_+g_ig^{-1}_i)=F^{(i)}_{+-}\,,\\
&D_+(g^{-1}_iD_-g_i)=F^{(i+1)}_{+-}\, ,
\end{split}
\end{equation}
which are in fact equivalent and where $F^{(i)}_{+-}=\partial_+A^{(i)}_--\partial_-A^{(i)}_+-\big[A^{(i)}_+,A^{(i)}_-\big]$.
 
\noindent Substituting the  constraints \eqref{5} into the equations of motion \eqref{6}  we find that
\begin{equation}\label{13}
\begin{split}
&\partial_+A^{(i)}_--\l^{-1}_i\l_0^{(i)}\partial_-A^{(i)}_+=\l^{-1}_i\l_0^{(i)}\big[A^{(i)}_+,A^{(i)}_-\big]\,,\\
&\l^{-1}_i(\l_0^{(i)})^{-1}\partial_+A^{(i)}_--\partial_-A^{(i)}_+=\l^{-1}_i(\l_0^{(i)})^{-1}\big[A^{(i)}_+,A^{(i)}_-\big]\,.
\end{split}
\end{equation}
After some algebra one can solve for the derivatives of the gauge fields to get
\begin{equation}\label{14}
\begin{split}
&\partial_-A^{(i)}_+=-\frac{1-(\l^{(i)}_0)^{-1}\l_i}{1-\l^2_i}\big[A^{(i)}_+,A^{(i)}_-\big]\,,\\
&\partial_+A^{(i)}_-=\frac{1-\l^{(i)}_0\l_i}{1-\l^2_i}\big[A^{(i)}_+,A^{(i)}_-\big]\,.
\end{split}
\end{equation}
In general, the Lax pair satisfies the Lax equation
\begin{equation}\label{15}
\partial_+\mathcal{L}_-(z)-\partial_-\mathcal{L}_+(z)=\big[\mathcal{L}_+(z),\mathcal{L}_-(z)\big],\quad \forall z\in\mathbb{C}\,.
\end{equation}
It is quite straightforward to see that in our case equations \eqref{14} are equivalent to \eqref{15} for
\begin{equation}
\mathcal{L}^{(i)}_{\pm}=\frac{2z}{z\mp 1}\tilde{A}^{(i)}_{\pm}\, , \quad \tilde{A}^{(i)}_{+}=\frac{1-\l^{(i)}_0\l_i}{1-\l^2_i}A^{(i)}_+\, ,\quad A^{(i)}_-=\frac{1-(\l^{(i)}_0)^{-1}\l_i}{1-\l^2_i}A^{(i)}_-\, ,
\end{equation}
where $z\in\mathbb{C}$ is the spectral parameter.

\section{Hamiltonian formulation}
In this appendix we will present the Hamiltonian formulation of  \eqref{d} for $n=2$,  but it can be easily generalized to arbitrary $n$,
with the goal to show that the Poisson Brackets of the two fields $B_+$, $A_-$ evaluated at the fixed point $(0)$ reduce to two independent equal time Kac-Moody algebras with central extension $k_2-k_1$. Let us remind to the reader that at the fixed point $(0)$ the two fields generate the chiral algebra symmetry of the corresponding CFT.

\no
Choosing a set of coordinates to parametrize the group valued fields $g_i$ we find that the action \eqref{d} for $n=2$  
in our conventions reads \footnote{In our conventions
	\begin{equation}
	\begin{split}
	&	\s^\pm = \tau \pm \s\, ,\quad  d\s^+\wedge  d \s^- = -2d^2 \s \, ,\quad d^2\s = d\tau \wedge d\s \ ,
	\\
	&
	\partial_+gg^{-1}=\frac{i}{2}R^a_{\m}(\dot{x}^{\mu}+x'^{\mu})t^a,\quad g^{-1}\partial_-g=\frac{i}{2}L^a_{\mu}(\dot{x}^{\mu}-x'^{\mu})t^a
	\end{split}
	\end{equation}
	}
\begin{equation}
\begin{split}
S&=\frac{k_1}{4\pi}\int d^2\s\Big(\frac{1}{2}R^a_{1\mu}R^{a}_{1\nu}(\dot{x_1}^{\mu}\dot{x_1}^{\nu}-x_1'^{\mu}x_1'^{\nu})+\l^{(1)}_{\m\n}\dot{x_1}^{\mu}x_1'^{\nu}\Big)+\\
&+\frac{k_2}{4\pi}\int d^2\s\Big(\frac{1}{2}R^a_{2\mu}R^{a}_{2\nu}(\dot{x_2}^{\mu}\dot{x_2}^{\nu}-x_2'^{\mu}x_2'^{\nu})+\l^{(2)}_{\m\n}\dot{x_2}^{\mu}x_2'^{\nu}\Big)+\\
&+\frac{k_1}{4\pi}\int d^2\s\Big(2iA^a_-R^{a}_{1\mu}(\dot{x_1}^{\mu}+x_1'^{\mu})-2iB^a_+L^a_{1\mu}(\dot{x_1}^{\mu}-x_1'^{\mu})+4B^a_+D^{ba}_1A^b_-\Big)\\
&+\frac{k_2}{4\pi}\int d^2\s\Big(2iB^a_-R^{a}_{2\mu}(\dot{x_2}^{\mu}+x_2'^{\mu})-2iA^a_+L^a_{2\mu}(\dot{x_2}^{\mu}-x_2'^{\mu})+4A^a_+D^{ba}_1B^b_-\Big)\\
&-\frac{\sqrt{k_1k_2}}{\pi}\int d^2\s\Tr\{B^a_+(\l^{-1}_2)^{ab}B^b_-+A^a_+(\l^{-1}_1)^{ab}A^b_-\}\,.
\end{split}
\end{equation}
The momenta canonically conjugate to the variables $(x_1, x_2, A_{\pm}, B_{\pm})$ are
\begin{equation}
\begin{split}
&\p_{\mu}^{(1)}=\frac{k_1}{4\pi}(R^a_{1\mu}R^a_{1\nu}\dot{x}^{\nu}+\l^{(1)}_{\mu\nu}x'^{\nu}+2iA^a_-R^a_{1\mu}-2iB^a_+L^a_{1\mu})\,,\\
&\p_{\mu}^{(2)}=\frac{k_1}{4\pi}(R^a_{2\mu}R^a_{2\nu}\dot{x}^{\nu}+\l^{(2)}_{\mu\nu}x'^{\nu}+2iB^a_-R^a_{1\mu}-2iA^a_+L^a_{1\mu})\,,\\
&P_{\pm}=\frac{\delta\mathcal{L}}{\delta\dot{A_{\pm}}}=0,\quad Q_{\pm}=\frac{\delta\mathcal{L}}{\delta\dot{B_{\pm}}}=0\,.
\end{split}
\end{equation}
Following \cite{5} and \cite{Hollowood:2014rla}  we define two sets of currents $J_{1\pm}$, $J_{2\pm}$ that obey two commuting Kac-Moody algebras at levels $k_1$, $k_2$ respectively
\begin{equation}
\Big\{J^a_{i\pm},J^b_{i\pm}\Big\}=\pm\frac{2\pi}{k_i}f^{abc}J^c_{i\pm}\delta_{\s\s'}\pm\frac{1}{k_i}\delta^{ab}\delta_{\s\s'}\,,
\end{equation}
and are given in terms of the initial variables as
\begin{equation}\label{B5}
\begin{split}
&J^a_{1+}=\frac{1}{2}R^a_{1\mu}(\dot{x}^{\mu}_1+ x'^{\mu}_1)+iA^a_{-}-iD^{ab}_1B^b_+\,,\\
&J^a_{1-}=\frac{1}{2}L^a_{1\mu}(\dot{x}^{\mu}_1- x'^{\mu}_1)-iB^a_{+}+iD^{ba}_1A^b_-\,.
\end{split}
\end{equation}
and
\begin{equation}\label{B6}
\begin{split}
&J^a_{2+}=\frac{1}{2}R^a_{2\mu}(\dot{x}^{\mu}_2+ x'^{\mu}_2)+iB^a_{-}-iD^{ab}_2A^b_+\,,\\
&J^a_{2-}=\frac{1}{2}L^a_{2\mu}(\dot{x}^{\mu}_1- x'^{\mu}_1)-iA^a_{+}+iD^{ba}_2B^b_-\,.
\end{split}
\end{equation}
We can show that the Hamiltonian can be rewritten in terms of $(J^{(1)}_{\pm}, J^{(2)}_{\pm},A_{\pm},B_{\pm})$ as
\begin{equation}
\begin{split}
H&=\frac{k_1}{4\pi}(J^a_{1+}J^a_{1+}+J^a_{1-}J^a_{1-}-4iA^a_-J^a_{1+}+4iB^a_+J^a_{1-}-2B^a_+B^a_+-2A^a_-A^a_-)\\
&+\frac{k_2}{4\pi}(J^a_{2+}J^a_{2+}+J^a_{2-}J^a_{2-}-4iB^a_-J^a_{2+}+4iA^a_+J^a_{2-}-2A^a_+A^a_+-2B^a_-B^a_-)+\\
&+\frac{\sqrt{k_1k_2}}{\pi}(B^a_+(\l_2^{-1})^{ab}B^b_-+A^a_+(\l_1^{-1})^{ab}A^b_-)\, .
\end{split}
\end{equation}
Thanks to Dirac we know that the primary constraints, i.e in our case $P(Q)_{\pm}=0$, can in principle generate secondary constraints from the demand that their time evolution on the constraint surface should vanish. In our case we obtain the following set of constraints
\begin{equation}\label{13'}
\begin{split}
&\Big\{P^a_{+},H\Big\}=0\quad \Rightarrow\quad  iJ^a_{2-}-A^a_++\l_0(\l^{-1}_1)^{ab}A^b_-=0\,,\\
&\Big\{P^a_{-},H\Big\}=0\quad \Rightarrow\quad iJ^a_{1+}+A^a_--\l_0^{-1}(\l^{-1}_1)^{ba}A^b_+=0\,,\\
&\Big\{Q^a_{+},H\Big\}=0\quad \Rightarrow\quad  iJ^a_{1-}-B^a_++\l_0^{-1}(\l^{-1}_2)^{ab}B^b_-=0\,,\\
&\Big\{Q^a_{-},H\Big\}=0\quad \Rightarrow\quad iJ^a_{2+}+B^a_--\l_0(\l^{-1}_2)^{ba}B^b_+=0\,.
\end{split}
\end{equation}
The set of all constraints, primary and secondary, are all second class which means that we can impose them strongly. Thus making use of \eqref{13'} the Hamiltonian takes the form
\begin{equation}
\begin{split}
H=&\frac{1}{4\pi}\bigg(\frac{k_2(\l^2_1-1)}{\l^2_1}\Tr A_+A_++\frac{k_1(\l^2_1-1)}{\l^2_1}\Tr A_-A_-\\
&+\frac{k_1(\l^2_2-1)}{\l^2_2}\Tr B_+B_++\frac{k_2(\l^2_2-1)}{\l^2_2}\Tr B_-B_-\bigg)\,,
\end{split}
\end{equation}
while the PBs of the fields $A_{\pm}$, $B_{\pm}$ are given by \footnote{ In what follows, we have calculated the Poisson brackets of the fields $A_\pm$ and $B_\pm$. 
Notice that for this particular choice of fields the Dirac and Poisson brackets are the same due to the protection mechanism discussed in  \cite{Hollowood:2014rla}.}
\ba
&&\Big\{A^a_+(\s),A^b_+(\s')\Big\}=-\frac{2\pi i }{k_2}f^{abc}\Big(f(\l_1,\l_0)A^c_-+g(\l_1,\l_0)A^c_+\Big)\delta_{\s\s'}+\frac{h(\l_1)}{k_2}\delta^{ab}\delta'_{\s\s'}\,,
\nonumber
\\
&&\Big\{A^a_-(\s),A^b_-(\s')\Big\}=-\frac{2\pi i }{k_1}f^{abc}\Big(g(\l_1,\l^{-1}_0)A^c_-+f(\l_1,\l^{-1}_0)A^c_+\Big)\delta_{\s\s'}-\frac{h(\l_1)}{k_1}\delta^{ab}\delta'_{\s\s'}\,,
\nonumber
\\
&&\Big\{A^a_+(\s),A^b_-(\s')\Big\}=-\frac{2\pi i}{k_1k_2}f^{abc}\Big(a(\l_1,\l_0)A^c_++b(\l_1,\l_0)A^c_-\Big)\delta_{\s\s'}\,,
\ea
and
\ba
&&\Big\{B^a_+(\s),B^b_+(\s')\Big\}=-\frac{2\pi i}{k_1}\Big(f(\l_2,\l_0^{-1})B^c_-+g(\l_2,\l_0^{-1})B^c_+\Big)\delta_{\s\s'}+\frac{h(\l_2)}{k_1}\delta^{ab}\delta'_{\s\s'}\,,
\nonumber
\\
&&\Big\{B^a_-(\s),B^b_-(\s')\Big\}=-\frac{2\pi i}{k_2}\Big(g(\l_2,\l_0)B^c_-+f(\l_2,\l_0)B^c_+\Big)\delta_{\s\s'}-\frac{h(\l_2)}{k_2}\delta^{ab}\delta'_{\s\s'}\,,
\nonumber
\\
&&\Big\{B^a_+(\s),B^b_-(\s')\Big\}=-\frac{2\pi i}{k_1k_2}f^{abc}\Big(b(\l_2,\l_0)B^c_++a(\l_2,\l_0)B^c_-\Big)\delta_{\s\s'}\,,
\ea
where we have made the following definitions\par
\vspace{2mm}
\begin{tabular}{ll}
\centering
		$f(\l_i,\l_0)=\displaystyle{\frac{h(\l_i)}{\l_i^2-1}(1-\l_0\l_i)}\,,$ & $g(\l_i,\l_0)=\displaystyle{\frac{h(\l_i)}{\l_i(\l_i^2-1)}}(\l_i^3-\l^{-1}_0)\,,$     \\
		$a(\l_i,\l_0)=\displaystyle{\frac{h(\l_i)}{\l_i^2-1}}(-k_2+k_1\l_i\l_0^{-1})\,,$& $b(\l_i,\l_0)=\displaystyle{\frac{h(\l_i)}{\l_i^2-1}}(-k_1+k_2\l_i\l_0)\,,$    \\
		$h(\l_i)=\displaystyle{\frac{\l_i^2}{\l_i^2-1}}\,.$
	\end{tabular}\par
\vspace{2mm}
\noindent Notice that in the case of equal levels, i.e. $\l_0=1$, the above PBs reduce to two copies of the $\l$-deformed algebra with two different deformation parameters. It is immediate to see that at the fixed point $(\l_1,\l_2)=(\l_0,\l_0)$ the above Poisson brackets simplify to
\begin{equation}
\label{B12}
\begin{split}
&\{A^a_+(\s),A^b_+(\s')\}=\frac{2\pi i}{k_2-k_1}f^{abc}\Big((1+\l^2_0)A^c_+-\l^2_0A^c_-)\delta_{\s\s'}+\frac{2\pi\l^2_0}{k_1-k_2}\delta^{ab}\delta'_{\s\s'}\,,\\
&\{A^a_-(\s),A^b_-(\s')\}=\frac{2\pi i}{k_2-k_1}f^{abc}A^c_-\delta_{\s\s'}+\frac{2\pi}{k_2-k_1}\delta^{ab}\delta'_{\s\s'}\,,\\
&\{A^a_+(\s),A^b_-(\s')\}=\frac{2\pi i}{k_2-k_1}f^{abc}A^c_+\delta_{\s\s'}\,.
\end{split}
\end{equation}
and
\begin{equation}
\label{B13}
\begin{split}
&\{B^a_+(\s),B^b_+(\s')\}=\frac{2\pi i}{k_2-k_1}f^{abc}B^c_+\delta_{\s\s'}-\frac{2\pi}{k_2-k_1}\delta^{ab}\delta'_{\s\s'}\,,\\
&\{B^a_-(\s),B^b_-(\s')\}=\frac{2\pi i}{k_2-k_1}f^{abc}\Big((1+\l_0^2)B^c_--\l^2_0B^c_+\Big)\delta_{\s\s'}-\frac{2\pi\l^2_0}{k_1-k_2}\delta^{ab}\delta'_{\s\s'}\,,\\
&\{B^a_+(\s),B^b_-(\s')\}=\frac{2\pi i}{k_2-k_1}f^{abc}B^c_-\delta_{\s\s'}\,.
\end{split}
\end{equation}
Notice that the fields $A_-, B_+$ form two commuting Kac-Moody algebras with central extension $k_2-k_1$.

\end{document}